\documentclass[useAMS,usenatbib]{mn2e}
\usepackage{graphicx}
\usepackage{amsmath}
\title[Theoretical line profiles of Hydrogen perturbed by collisions with protons]{Theoretical study of the line profiles of the hydrogen perturbed by collisions with protons}

\author[M. G. Santos and S. O. Kepler]{M. G. Santos$^{1}$\thanks{E-mail:
marcios@if.ufrgs.br} and S. O. Kepler $^{1}$\\
$^{1}$Instituto de F\'{\i}sica, Universidade Federal do Rio Grande do Sul, 91501-900 Porto Alegre, RS, Brazil}

\begin{document}

\date{Accepted 2012 January 24. Received 2012 January 17; in original form 2011 September 22}

\pagerange{\pageref{firstpage}--\pageref{lastpage}} \pubyear{2011}

\maketitle

\label{firstpage}

\begin{abstract}

We present theoretical calculations of the quasi-molecular line profiles for the Lyman (Ly$\alpha$, Ly$\beta$, Ly$\gamma$, Ly$\delta$) and Balmer (H$\alpha$, H$\beta$, H$\gamma$, H$\delta$, H$\epsilon$, H$8$, H$9$, and H$10$) series,  perturbed by collisions with protons.  In all calculations we include the dependence of the dipole moments as a function of internuclear distance during the collision.  The broadening from ion collisions must be added to the normal electron Stark broadening.
\end{abstract}


\section{Introduction}

The satellite lines present in the far wing profiles of the Lyman and Balmer series lines of atomic hydrogen have been observed in stars and in laboratory plasmas \citep{Kielkopf2002}.  Experimentally, the primary difficulties in studying these line profiles are (1) the development of techniques to observe neutral atomic hydrogen interactions at densities high enough for spectral line broadening effects to be significant, and (2) quantitatively characterizing the conditions under which the spectra are created.  On the theoretical side, the difficulty arives in developing methods for computing the contributions from all possible molecular states of H$_{2}$ and H$_{2}^{+}$ to the spectrum of free atoms in collision.

Satellite features at 1600~\AA~ and 1405~\AA~ in the Lyman$-\alpha$ wing associated with free-free quasi-molecular transitions of H$_{2} $ and H$_{2}^{+}$ are observed in the ultraviolet (UV) spectra of a few stars obtained with the International Ultraviolet Explorer (IUE) and the Hubble Space Telescope (HST).
The satellites in the red wing of Lyman$-\beta$ are in the 905 to 1187~\AA~ spectral region, covered by Far Ultraviolet Espectroscopic Explorer (FUSE).
The stars which show Lyman satellites are DA white dwarfs (WDs), old Horizontal Branch stars of spectral type A, and $\lambda$ Bootis stars \citep{Allard2000}.  WDs are the end product of evolution of $\sim 98$\% of all stars \citep[e.g.,][]{Kepler2007}.  Having consumed all available nuclear fuel (which depends on the individual stellar mass), their source of luminosity is dominated by only the residual internal thermal energy.  This radiates away, exponentially cooling the star, but due to the small surface area (a typical WD radius is $\sim$10$^9$~cm), cooling times are long, taking approximately 10$^{10}$~years for their effective temperatures to decrease from around 10$^5$~K to near 10$^3$~K.  Therefore, WDs are old, and hidden in their structure is a historical record of stellar formation and evolution in our Galaxy.

UV spectroscopic observations of WDs reveal a line shape very different from the expected simple Stark broadening, with satellite lines in the red wing of Lyman$-\beta$ near 1078 and 1060~\AA~ \citep{Koester1996,Koester1998}.   The strengths of these satellite features, and indeed the entire shape of the wings in the Lyman series, are very sensitive to the degree of ionization in the stellar atmosphere or laboratory plasma, because it is the ionization that determines the relative importance of broadening by ion and neutral collisions.  Therefore the satellites may be used as a temperature diagnostic in a plasma where Saha ionization-equilibrium holds.  In the case of Lyman$-\alpha$ and the H - H$^{+}$ Lyman$-\beta$ satellites, the shape of the potential plays a dominant role in the large difference in the broadening of quasi-molecular features \citep{Allard1998b}.

Quasi-molecular transitions occur when a photon is absorbed or emitted by a hydrogen atom while this atom interacts with one or more neighboring particles, which may be atoms, ions or molecules \citep{Allard1982}.  At short distances, the interaction between the perturbed atoms and the other particle(s) usually leads to the formation of satellite molecular lines on the wings of atomic lines \citep{Allard1998a,Allard1998b,Allard1999}.  Structures observed in the wing of Lyman$-\alpha$ at 1623~\AA~ and 1405~\AA~ are due to absorptions of quasi-molecular H$_{2}$ and H$_{2}^{+}$, respectively.  The intensity of these two satellites depends strongly on the degree of ionization in the stellar atmosphere and consequently on the effective temperature T$_\mathrm{eff}$ and surface gravity $\log g$ \citep{Holweger1994}.

The atmosphere of a WD comprises less than $10^{-13}$ of the stellar mass and is dominated by hydrogen (DA white dwarfs) or helium (DB/DO) and sometimes shows small amounts of heavy elements (e.g., DAZ, DBZ and DOZ white dwarfs) \citep[see, e.g.,][]{Koester2010}.  The primary technique to determine T$_\mathrm{eff}$ and $\log g$ of WDs is by comparing observed spectra with synthetic spectra from theoretical atmosphere models \citep[e.g.,][]{Tremblay2009}.  The advantage of this spectroscopic technique is that the theoretical line profiles are extremely sensitive to variations in the atmospheric parameters.  \citet{Bergeron1992} were one of the first to apply this to a large number of WDs; they fit the hydrogen Balmer lines.  The technique was later applied to the UV Lyman lines \citep[e.g.,][]{bergeron95, Kepler1993, Vennes2005}.

\citet{Bergeron2002} made improvements to the models by taking into account the optical quasi-molecular opacities due to the dipole moments induced by the interaction between particles.  This interaction causes a temporary rearrangement of electronic charges, which forms a temporary molecule, thus producing absorptions or emissions beyond those of the isolated atoms.

Currently there are a large number of optical spectra of WDs that have been taken as part of the Sloan Digital Sky Survey (SDSS) \citep[e.g.,][]{Eisenstein2006,Kleinman10}, allowing for the study of the WD mass distribution \citep{Kepler2007}.  \citet{Kleinman2004} and \citet{Liebert2005} derive masses from spectroscopic fits, and \citet{Bergeron1991} and \citet{Koester1991} discuss the apparent increase in the average mass of DAs with T$_\mathrm{eff}<$12\,000~K.  \citet{Kepler2006} compare SDSS spectra with high S/N spectra obtained with the 8~m Gemini telescope for four WDs with T$_\mathrm{eff}\sim$ 12\,000~K and determine that the masses derived from fitting the relatively low S/N SDSS spectra are systematically overestimated by $ \Delta M\simeq 0.13  M_\odot $.  Due to a correlation between $T_\mathrm{eff} $ and $\log g$ (a small increase in $T_\mathrm{eff} $ can be offset by a small decrease in $\log g$), such discrepancies are concentrated only in the region around the maximum of the Balmer lines, 14\,000~K~$\geq T_\mathrm{eff} \geq$~12\,000~K.  To study the trend of the apparent increased average mass of DAs, \citet{Kepler2010} analyse 1505 such WDs with $S/N\geq 20$ and T$_\mathrm{eff}>12\,000$~K and determined an average DA mass of $ M / M_\odot=0.604\pm 0.003$.  They observe that the distribution of the sample is similar to that of the Palomar Green survey published by \citet{Liebert2005}.
\citet{Falcon2010} used an ensemble average of gravitational redshifts of 449 DAs, which is independent of line profiles, and estimated a mean mass of $0.640\pm 0.014M_\odot$. 

\citet{Koester09}, \citet{Koester2010}, \citet{Tremblay2009} and \citet{Gianninas2010} study the possible causes for this apparent increase in mass, not seen in photometric determinations \citep{Koester07, Kepler2007}.  They conclude this phenomenon is not caused by He contamination, but \citet{Tremblay2011} propose that the use of one-dimensional calculations of convection in the atmosphere, compared to more realistic three-dimensional calculations, might be the cause.  There is a huge effort to eliminate the inconsistencies in the model calculations, and these attempts show the need for continued study of the interaction of hydrogen perturbed by protons, electrons, atoms and molecules at high pressure.

The goal of our work is the study of the interaction of the hydrogen atom with protons, which temporarily creates the molecule H$_{2}^{+}$.  We calculate the potential of H$_{2}^{+}$ up to $n=10$, and analyze all possible transitions that can occur, as well as the dipole moments for all possible transitions.  In the literature, we find calculations for only Ly$\alpha$, Ly$\beta$ and Ly$\gamma$ and H$\alpha$; for these, we calculate to increased resolution.  This new theoretical data must be inserted in the modeling of line profiles at high density, in addition to the electron Stark broadening.  

\section{Theoretical analysis}

The dipole moments induced by the interaction between radiator and perturber cause a temporary repositioning of electronic charges, which temporarily forms a molecule.  This leads to the possible appearance of satellite features in the wing of the line profile and produces different absorptions or emissions than those of non-interacting atoms \citep{Lothar1993}.

The position of the satellite line depends directly on the extreme of the potential difference of the transition $\Delta V_{ext}(R)$ and the dependence of the potential difference $\Delta V(R)=V_{j}(R)-V_{i}(R)$ with internuclear separation $R$ of the states $ i $ and $ j $ of the allowed transitions of the H$_{2}^{+}$; the intensity and shape of the satellites depend on both the potential difference and the radiative dipole moment \citep{Allard2004}.

To calculate spectral line profiles, affected by collisions with protons,
we must know precisely the theoretical potential that describes the molecular interaction between radiator and perturber, as well as the changes in dipole momenta due to the radioactive atom-ion separation for each molecular state.

\citet{Madsen1971} calculate the values of the potential of H$_{2}^{+}$ only for states with $n\leq  3$.  \citet{Allard2004} calculate the potential for states with $n\leq 4$, which allows them to calculate the line profile for Lyman$-\gamma$.

The values of the dipole moments of H$_{2}^{+}$ that are available in the literature were calculated by \citet{Ramaker1972, Ramaker1973}, and \citet{Allard2009} refers to the calculation of $n=4$.

For this paper, we calculate all potential and dipole moments up to $n\leq 10$ of H$_{2}^{+}$ as a function of the internuclear distance $R$ up to 300 a.u. (atomic units), with a step of 0.01 a.u.  We compare the potential of the n~$\leq$~3 states with those in the literature \citep{Madsen1971} and show differences smaller than $10^{-13}$, the precision quoted in the published values.  We also compare the computed dipole moments with those tabulated by \citet{Ramaker1973} and those plotted in \citet{Allard2009}; they are fully consistent.

\subsection{Potentials for the Hydrogen Molecule H$_{2}^{+}$}
 
The structure of the molecular ion H$_{2}^{+}$ has been studied by \citet{Teller1930}, \citet{Hylleraas1931}, \citet{Jaffe1934}, \citet{Baber1935}, \citet{Bates1953}, and \citet{Madsen1971}, but accurate tables of the electronic energy as a function of internuclear distance for the excited states n~$\geq$~3 have not been published.

As originally shown by \citet{Burrau1927}, the Schr\"odinger equation for 
H$_{2}^{+}$ can be separated and the electronic energy $E$ calculated exactly.
The Schr\"odinger equation of the electron in atomic units is:
\begin{equation}
\label{schrodinger}
\frac{1}{2}\nabla^{2}\Psi+\Bigl(E+\frac{1}{r_{a}}+\frac{1}{r_{b}}-\frac{1}{R}\Bigr)\Psi=0
\end{equation}
where $r_{a}$ and $r_{b}$ are the distances measured in units of $a_{0}$ of the electron to the two positive charges $a$ and $b$, $E$ is the energy measured in units of Hartree $E_{h}$, and $R$ is the internuclear separation.

In terms of confocal elliptical coordinates, defined by:
$\lambda=({r_{a}+r_{b}})/{R}$,
$\mu=({r_{a}-r_{b}})/{R}$
and
$\phi$, the azimuthal angle with respect to the internuclear axis,
the Schr\"odinger equation can then be separated in three equations:
\begin{equation}
\label{eqphi}
\frac{\partial^{2}\Phi}{\partial\phi^{2}}+m^{2}\Phi=0 
\end{equation}

\begin{equation}
\label{eqmi}
\frac{\partial}{\partial\mu}\left[ (1-\mu^{2})\frac{\partial M}{\partial\mu}\right]+\left(-A+p^{2}\mu^{2}-\frac{m^{2}}{1-\mu^{2}}\right)M=0
\end{equation}
and
\begin{equation}
\label{eqlambda}
\frac{\partial}{\partial\lambda}\left[ (\lambda^{2}-1)\frac{\partial \Lambda}{\partial\lambda}\right]+\left(A+2R\lambda - p^{2}\lambda^{2}-\frac{m^{2}}{\lambda^{2}-1}\right)\Lambda=0 
\end{equation}
where $\Psi(\lambda,\mu,\phi)=\Lambda(\lambda)M(\mu)\Phi(\phi)$, $p^{2}=\frac{-\varepsilon R^{2}}{2}$, 
$\varepsilon=E-\frac{1}{R}$ is the electronic energy, and where $A$ and $m$ are the constants of separation.

The solution of equation (\ref{eqphi}) is direct, i. e.:
\begin{equation}
\label{solphi}
\Phi=e^{(\pm im\phi)}.
\end{equation}
The solution of equation (\ref{eqmi}), the angular component, is obtained using the \citet{Hylleraas1931} method:
\begin{equation}
\label{solm}
M(\mu)=\sum_{s}c_{s}P_{m+s}^{m}(\mu),
\end{equation}
where $P_{m+s}^{m}(\mu)$ are the generalized Legendre polynomials of first order.
The summation extends only over even or odd values of $s$,
depending on the state calculated.

Applying equation (\ref{solm}) in equation (\ref{eqmi}), we obtain a recurrence relation for the coefficients $c_{s}$,
\begin{equation}
\label{recmi}
a_{s}c_{s-2}+b_{s}c_{s}+d_{s}c_{s+2}=0
\end{equation}
where:

$
a_{s}=p^{2}\frac{(s-1)s}{(2s+2m-3)(2s+2m-1)}
$
\\

$
b_{s}=-A-(s+m)(s+m+1)+
$
\\

$
\quad\quad +p^{2}\left[\frac{(s+1)(s+1+2m)}{(2s+2m+3)(2s+2m+1)}+\frac{s(s+2m)}{(2s+2m+1)(2s+2m-1)}\right]
$
\\ 

$
d_{s}=p^{2}\frac{(s+2m+2)(s+2m+1)}{(2s+2m+5)(2s+2m+3)}
$
\\

The recurrence relation (\ref{recmi}) forms an infinite set of homogeneous linear equations whose solution requires the determinant of the coefficient matrix to be null.

The solution of equation (\ref{eqlambda}) for the radial part is obtained using the method of \citet{Jaffe1934}.
\begin{equation}
\label{sollamb}
\Lambda(\lambda)=(\lambda^{2}-1)^{\frac{m}{2}}(1+\lambda)^{\sigma}e^{-p\lambda}\sum_{t=0}^{\infty}g_{t}\left(\frac{\lambda-1}{\lambda+1}\right)^{t} ,
\end{equation}
where $ \sigma =\frac{R}{p}-m-1 $.

Applying equation (\ref{sollamb}) to the equation (\ref{eqlambda}), the following recurrence equation is found:
\begin{equation}
\label{reclambda}
\alpha_{t}g_{t+1}-\beta_{t}g_{t}+\gamma_{t}g_{t-1}=0.
\end{equation}
where: 
\\
$
\alpha_{t}=(t+1)(t+m+1),
$
\\
$
\beta_{t}=2t^{2}+(4p-2\sigma)t-A+p^{2}-2p\sigma-(m+1)(m+\sigma),
$
\\
$
\gamma_{t}=(t-1-\sigma)(t-1-\sigma - m).
$
\\

The recurrence relation (\ref{reclambda}) also produces an infinite set of homogeneous linear equations, whose determinant of the matrix of coefficients must also be null.
When calculating the determinant of the infinite recurrence relations of equations (\ref{recmi}) and (\ref{reclambda}),
we obtain two functions $f_{1} (A, \varepsilon, R)$ 
and $f_{2} (A,\varepsilon, R)$ which depend on $A$, $\varepsilon$, and $R$. 
The electronic energy (potential energy $V_{i}$) $\varepsilon $ is obtained by calculating the solution 
of both functions $f_{1} (A, \varepsilon, R)$ and $f_{2} (A, \varepsilon, R)$, 
e.g., by using the Newton-Raphson method.  Practical considerations require the truncation of the recurrence relations (\ref{reclambda}) and (\ref{recmi}).  We choose t=s=25 for 
the data presented here, because larger values for $t$ and $s$ do not produce any 
significant change in $A$ and $\varepsilon$.
The coefficients $c_{s}$ and $g_{t}$ are calculated using the method of continued fractions.

\subsection{Electric dipole moments}
Choosing the origin of the coordinate system in the center of the line segment which unites the two positive charges and taking this line
as the axis $z$, the matrix element to be evaluated of the components of the electric dipole moment, $D_{ij} $, specified by:
\begin{equation}
\label{dip0}
D_{ij}=e\left\langle i |\textbf{r}| j \right\rangle =  e Q_{i,j}(\textbf{r}) \; , 
\end{equation}
where $e$ represents the electron charge, \textbf{r} is the 
position vector of the electron relative to molecular fixed frame, in coordinates $x$, $y$ and $z$ of the electron, $ \Psi_{i} $ and $ \Psi_{j} $ are the 
electronic functions in the Born-Oppenheimer approximation of the states $ i $ and $ j $ and $Q_{i,j}(\textbf{r})$ the matrix elements.

The matrix elements $Q_{i,j}(\textbf{r})$ of the various components of the electric dipole moment according to \citet{Ramaker1973}, 
are specified by:
\begin{equation}
\centering
\label{dip}
Q_{i,j}(\textbf{r})=\int \Psi^{*}_{i}(\lambda ,\mu ,\phi) \left( \textbf{r} \right) \Psi_{j} (\lambda ,\mu ,\phi) d\tau \;.
\end{equation}
where $\tau$ is the volume element. $Q_{i,j}(\textbf{r})$ is obtained by inserting the equations (\ref{solphi}), (\ref{solm}) and (\ref{sollamb}) into equation (\ref{dip}).
The matrix elements $ Q_{i,j}(\textbf{r}) $ can be written as a function of the coordinates $x$, $y$ and $z$ as follows:
\begin{eqnarray}
Q_{i,j}(x) =N_{i}N_{j}\left( \frac{R}{2}\right)^{4}
\int^{2\pi}_{0} \Phi^{*}_{i}(\phi) cos(\phi) \Phi_{j} (\phi) d\phi\nonumber \\
\times\left[\Omega_{i,j}\left(\frac{1}{2},0,\frac{1}{2},2\right) - \Omega_{i,j}\left(\frac{1}{2},2,\frac{1}{2},0 \right) \right]
\label{dipx}
\end{eqnarray}

\begin{eqnarray}
Q_{i,j}(y)=N_{i}N_{j}\left( \frac{R}{2}\right)^{4}
\int^{2\pi}_{0} \Phi^{*}_{i}(\phi) sin(\phi) \Phi_{j} (\phi) d\phi\nonumber \\
\times\left[\Omega_{i,j}\left(\frac{1}{2},0,\frac{1}{2},2\right) - \Omega_{i,j}\left(\frac{1}{2},2,\frac{1}{2},0 \right) \right]
\label{dipy}
\end{eqnarray}

\begin{eqnarray}
Q_{i,j}(z)=N_{i}N_{j}\left( \frac{R}{2}\right)^{4}
\int^{2\pi}_{0} \Phi^{*}_{i}(\phi) \Phi_{j} (\phi) d\phi\nonumber \\
\times\left[\Omega_{i,j}(0,1,0,3) - \Omega_{i,j}(0,3,0,1) \right]
\label{dipz}
\end{eqnarray}
where $ \Omega_{i,j}(a,b,c,d)$ is defined by:

\begin{eqnarray}
\Omega_{i,j}(a,b,c,d) =\int^{1}_{-1} M^{*}_{i}(\mu)(1-\mu^{2})^{a}\mu^{b}M_{j}(\mu) d\mu\nonumber \\
\times\int^{\infty}_{1} \Lambda^{*}_{i}(\lambda)(\lambda^{2}-1)^{c}\lambda^{d}\Lambda_{j}(\lambda) d\lambda
\label{omeg}
\end{eqnarray}

The normalization constants can be calculated from:
\begin{eqnarray}
N_{i}^{-2} = \left( \frac{R}{2}\right)^{3}  [\Omega_{i,i}(0,0,0,2) - \Omega_{i,i}(0,2,0,0) ] \nonumber \\
\times\int \Phi^{*}_{i}(\phi) \Phi_{i} (\phi) d\phi
\label{normaA}
\end{eqnarray}

\begin{eqnarray}
N_{j}^{-2} = \left( \frac{R}{2}\right)^{3} [\Omega_{j,j}(0,0,0,2) - \Omega_{j,j}(0,2,0,0) ]\nonumber \\
\times\int \Phi^{*}_{j}(\phi) \Phi_{j} (\phi) d\phi
\label{normaB}
\end{eqnarray}

After calculating the values ​​of $ Q_{i,j}(\textbf{r}) $ for each components x, y and z, we obtain $D_{ij} $ as follows \citep{Herman1956}:
\begin{equation}
\centering
|D_{ij}|^{2}=|eQ_{i,j}(x)|^{2}+|eQ_{i,j}(y)|^{2}+|eQ_{i,j}(z)|^{2} 
\label{dipquad}
\end{equation}

We compute the values of the electric dipole moments, equation (\ref{dipquad}), for all transitions of the Lyman and Balmer series for the H$_{2}^{+}$ molecular ion for internuclear separations up to 300 a.u. \;.

Table~\ref{dipcomp} shows values of the electric dipole moments of this work compared with those of \citet{Ramaker1973} for some transitions, demonstrating that our work presents a higher precision and greater coverage of the internuclear distance.

For equations (\ref{solm}) and (\ref{sollamb}) to be solutions of the appropriate electronic functions $\Psi_{i}$, the summations must be allowed to go to infinity.  However, we achieve convergence of the series for $s=31$ and $t=10$.

The data of the potential and electric dipole moments are plotted in subsequent sections for the Lyman series (Ly$\alpha$ to Ly$\delta$) and Balmer-$ \alpha $. Electric dipole moments are plotted for Balmer-$ \alpha $ and Balmer-$ \beta $.

\begin{table}
\centering 
  \caption{Table of comparison between the data of \citet{Ramaker1973}$^{\dagger}$ and this work$ ^{\ddagger} $ of the electric dipole moments for Lyman-$\alpha$ of the H$_{2}^{+}$ in atomic units.}
  
\begin{tabular}{|c |  c | c | c | c |} 
\hline
$R(a_{0})$ & Transition & $ Q_{\alpha,\beta}(R)^{\dagger} $ & $ Q_{\alpha,\beta}(R)^{\ddagger} $\\
\hline
1.0 & $2p\pi_{u}$-$1s\sigma_{g}$ & 0.5494 &  0.5493774 \\
11.0 & $2p\pi_{u}$-$1s\sigma_{g}$ & 0.7169 & 0.7168555 \\
60.0 & $2p\pi_{u}$-$1s\sigma_{g}$ & 0.7450 & 0.7449931 \\

1.0 & $3p\sigma_{u}$-$1s\sigma_{g}$ & 0.1651 &  0.1651107 \\
11.0 & $3p\sigma_{u}$-$1s\sigma_{g}$ & 0.4859 & 0.4858565 \\
80.0 & $3p\sigma_{u}$-$1s\sigma_{g}$ & 0.5207 & 0.5207119 \\

1.0 & $4f\sigma_{u}$-$1s\sigma_{g}$ & 0.001055 & 0.0010541 \\
11.0 & $4f\sigma_{u}$-$1s\sigma_{g}$ & 0.6609 &  0.6608712 \\
80.0 & $4f\sigma_{u}$-$1s\sigma_{g}$ & 0.5326 &  0.5325508 \\

1.0 & $2s\sigma_{g}$-$ 2p\sigma_{u}$ & 1.241 &   1.2406089 \\
11.0 & $2s\sigma_{g}$-$ 2p\sigma_{u}$ & 0.4844 & 0.4844248 \\
80.0 & $2s\sigma_{g}$-$ 2p\sigma_{u}$ & 0.5207 & 0.5207119 \\

1.0 & $3d\sigma_{g}$-$ 2p\sigma_{u} $ & 1.010 &   1.0104411 \\
11.0 & $3d\sigma_{g}$-$ 2p\sigma_{u} $ & 0.6349 & 0.6349347 \\
80.0 & $3d\sigma_{g}$-$ 2p\sigma_{u} $ & 0.5326 & 0.5325508 \\

1.0 & $3d\pi_{g}$-$ 2p\sigma_{u} $ & 0.8584 &  0.8584398 \\
11.0 & $3d\pi_{g}$-$ 2p\sigma_{u} $ & 0.7935 & 0.7935359 \\
60.0 & $3d\pi_{g}$-$ 2p\sigma_{u} $ & 0.7450 & 0.7449931 \\

\hline
\end{tabular} 
\label{dipcomp}
\end{table} 

\section{Profiles and Satellites}

Currently there are different approaches to the calculation of the line profiles, namely, quasi-static, quantum-mechanical and unified theory.

In the quasi-static approximation, the expression of the cross-section may be written as \citep{Margenau1959,Allard1982,Rohrmann2011}

\begin{equation}
\centering
\label{profile}
\sigma_{ij}(\nu)=\frac{16}{3}\pi^{3}n_{p}\alpha\biggl\{\frac{h\nu}{|dV_{ij}/dR|}\biggr\}D_{ij}R^{2}e^{(4 \pi n_{p} R^{3}/3)} e^{-\beta E_{i}(R)},
\end{equation}
where $\beta=\frac{1}{kT}$, k is the Boltzmann constant, and T is the gas temperature.  $n_{p}$ is the mean density of perturbers in the gas, $V_{ij}$ the potential energy difference associated with a transition with an energy $h\nu = V_{ij} \equiv V_{j}(R)-V_{i}(R)$ at internuclear separation $R$, $D_{ij}(R)$ is the dipole moment, and $E_{i}(R)$ corresponds to the change in energy (with respect to $R=\infty$) in the state $i$ of the H$_{2}^{+}$. When there is an extreme in the curve corresponding to the difference in the potential energies between the upper and lower states, $\frac{dV_{ij}}{dR}$ is zero.  Therefore, if the derivative is zero, there is a divergence, called a satellite line.

In the quantum-mechanical approach, the cross section for the spontaneous plus stimulated association process is given by 
\citep{Zygelman1990}

\begin{eqnarray}
\sigma(E)=\sum_{N^{'}}\sum_{v^{''}}\frac{64}{3}\frac{\pi^{5}}{c^{3}}\frac{\nu^{3}}{\kappa^{2}}\frac{1}{1-exp(-h\nu /kT)} \nonumber \\
\times p[N^{'}M^{2}_{v^{''},N^{'}-1;\kappa ,N^{'}} + (N^{'}+1)M^{2}_{v^{''},N^{'}+1;\kappa ,N^{'}}] \nonumber \\
\label{eqdalgarno}
\end{eqnarray}
where $E$ is the relative collision energy, $ \kappa $ is the wave number of relative motion, $p$ is the probability of approach in the initial electronic state, $N^{'}$ is the initial rotational quantum number, $ v^{''} $ is the final vibrational quantum number, and M is the electric dipole matrix element connecting the initial continuum and final rotational-vibrational states of the ground electronic state.

The unified theory has been developed in \citet{Allard1999}, and a detailed discussion is presented there.  The fundamental expression of the normalized spectrum line $ F_{\nu}(\Delta\nu) $ is computed from the Fourier transform \citep{Allard2009}

\begin{equation}
\centering
\label{eqallard}
F_{\nu}(\Delta\nu)=\frac{1}{\pi}Re\int_{0}^{+\infty}\Phi(s)e^{-i\Delta\nu s}ds.
\end{equation}

\begin{figure}
\centering 
\includegraphics[width=84 mm,clip=]{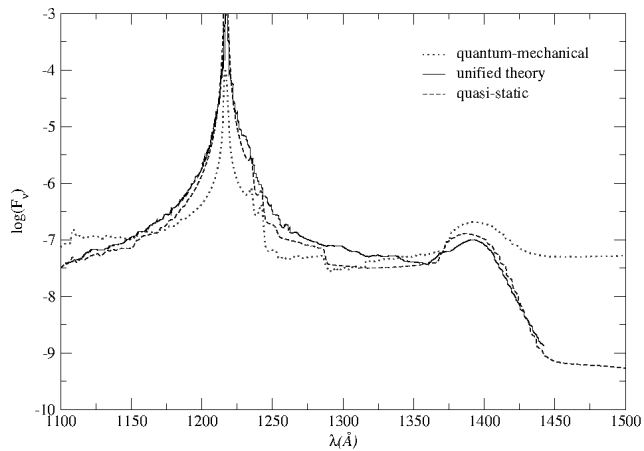}
\caption{Comparison of the line profile of Lyman-$\alpha$ normalized between the three approaches, quasi-static (dashed line), 
quantum-mechanical (dotted line) and unified theory (solid line) adapted from \citet{Allard1999}.}
\label{comply}
\end{figure}

Figure \ref{comply} shows the comparison of the line profile of Ly$\alpha$ of the three approaches mentioned above.  The solid line shows the approach from the unified theory, which has a resolution determined by Fourier transform (equation \ref{eqallard}), while the quantum-mechanical approaches (dashed line) and quasi-static (solid line) values provide similar but distinct profiles due to the calculation of the derivative by the numerical difference in the quasi-static and by the series expansion in quantum-mechanics.  The three profiles are similar; however, the quasi-static approach better represents the correct 
global shape of the line wings and position of the satellites, in comparison with previously published data \citep{Allard1999}.
As the main aim of our work are: the calculations of the electric dipole moments, the potentials of H$_{2}^{+}$ ion, the calculations of 
the global shape of line profiles and the location of the satellites, we use the quasi-static approach to calculate the line profiles.
Unfortunately, the quasi-static approach does not provide an appropriate shape of the center of the lines, thus there is a need for better 
modeling of the center lines of each transition.

Our interest is to find the correct position of the satellites in the wings of the line profile and their general form (i.e., the divergences in the profile) and bring these results to the attention of astrophysicists.  If they are observable, they will be specific for protons (although other ions may give qualitatively similar effects).

The method we used to obtain the profiles following \citet{Rohrmann2011}.
For lines profile of the Lyman and Balmer series in a quasi-static approach, we identified each pair of 
levels $(i,j)$ of the quasi-molecule H - H$^{+}$ 
that contributes to the formation of a given line. Only transitions between states u (ungerade) and g (gerade) are allowed, 
i.e., with opposite parity and taking 
$\mid{m-m'}\mid=0$ or 1 as the allowed dipole.
The second step is to calculate the effective cross section $\sigma_{ij}(\nu)$, equation (\ref{profile}) using the potentials 
and dipole moments 
for the transitions already identified. 
For the calculation of $\sigma_{ij}(\nu)$ we need to analyze the frequencies 
$\nu$ that arise from the difference in energy between these levels $h\nu = V_{j}-V_{i}$. 
However, we must take into account that 
different internuclear distances 
can produce the same frequency.
The effective cross section is the sum of each of these contributions.

The line profile is the sum of the effective cross sections $ \sigma_{ij}(\nu) $ multiplied by the statistical weights of each of the 
respective transitions.
The calculated profile was convolved with a Maxwellian distribution of 
velocities to take into account the thermal Doppler broadening.

The line profiles and satellite lines of Lyman and Balmer series shown in this paper are for
single densities of
$10^{16}$ to $10^{18}~\mathrm{cm}^{-3}$, typical of the atmosphere of white dwarf stars and that can be produced in z-pinch laboratories.
We remind the reader that the real profile must include the electrons Stark broadening, which do not have satellites.

\subsection{Lyman series}

\subsubsection{Lyman-$\alpha$}

\begin{figure}
\centering 
\includegraphics[width=84mm,clip=]{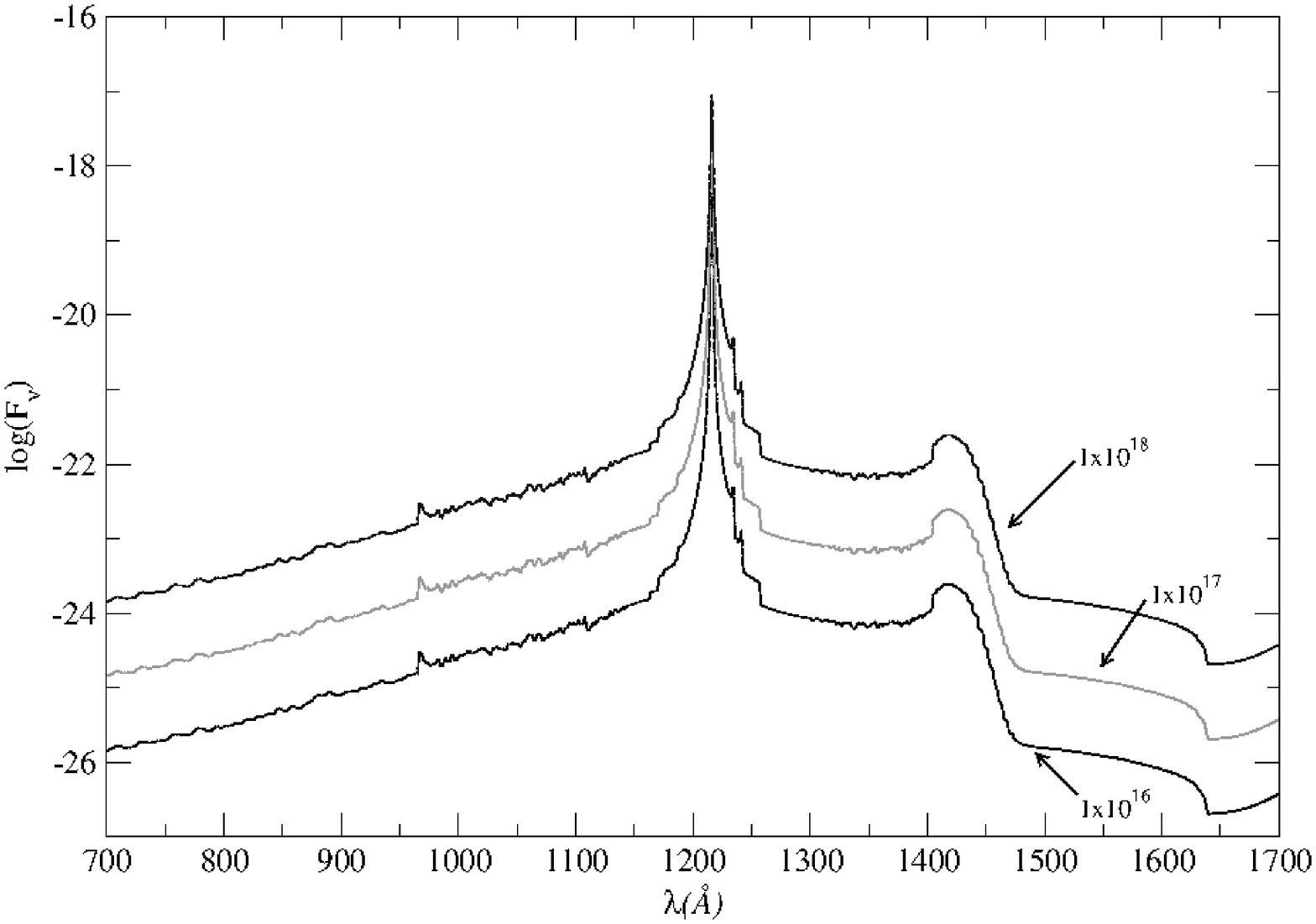}
\caption{Lyman-$\alpha$ profile with density of H$^{+}$ perturbers (T$=4\,000$~K), 
$n_{H}^{+}=10^{16}$, $n_{H}^{+}=10^{17}$ and $n_{H}^{+}=10^{18}$ cm$^{-3}$ (from bottom to top).}
\label{LyA13}
\end{figure}

\begin{figure}
\centering 
\includegraphics[width=84mm,clip=]{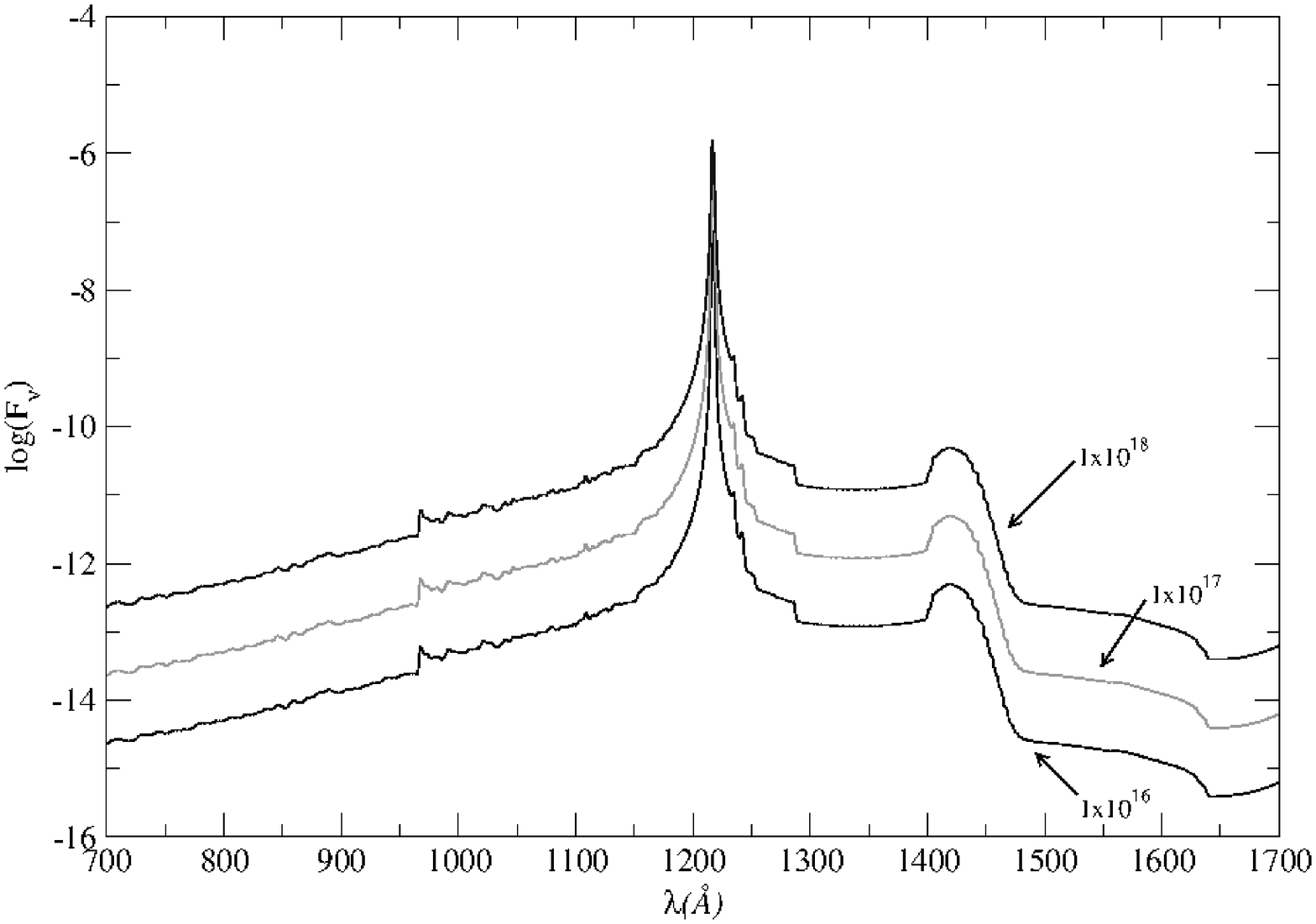}
\caption{Lyman-$\alpha$ profile with density of H$^{+}$ perturbers (T$=12\,000$~K), 
$n_{H}^{+}=10^{16}$, $n_{H}^{+}=10^{17}$ and $n_{H}^{+}=10^{18}$ cm$^{-3}$ (from bottom to top).}
\label{LyA1012}
\end{figure}

The Lyman-$\alpha$ total profile depends on 6 individual transitions. 
The line profile calculations shown in  Figures \ref{LyA13} and \ref{LyA1012} have been done for  temperatures of $4\,000$~K and $12\,000$~K, 
at proton densities of $10^{16}$, $10^{17}$ and $10^{18}$ cm$^{-3}$.

\begin{figure}
\centering 
\includegraphics[width=84mm,clip=]{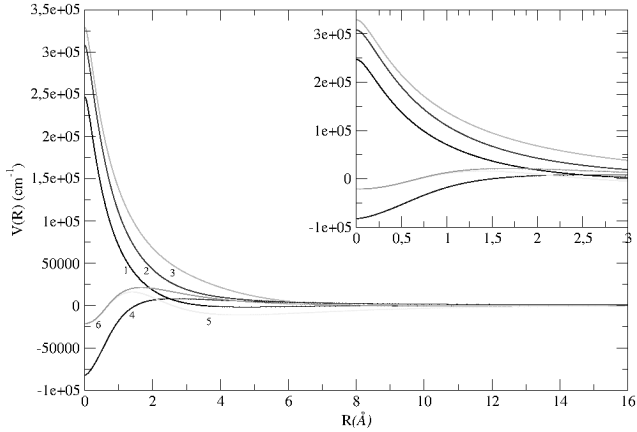}
\caption{Transitions difference potential as a function of interatomic distance of the H$_{2}^{+}$ of Lyman-$\alpha$.}
\label{Ly2all}
\end{figure}

\begin{figure}
\centering 
\includegraphics[width=84mm,clip=]{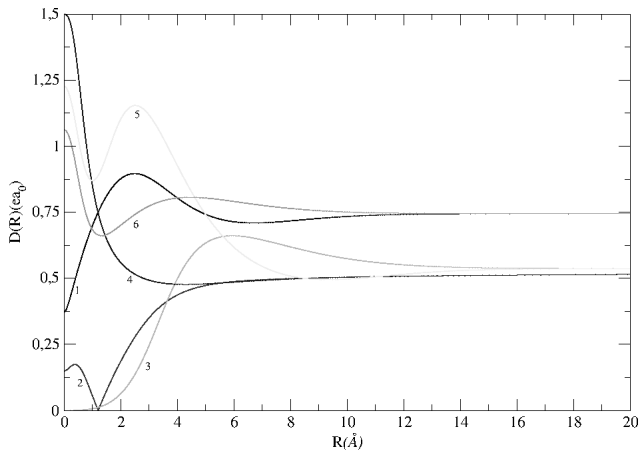}
\caption{Electronic transition dipole moments of Lyman-$\alpha$ due to H perturbed by collisions with protons. The $D(R)$ is given here 
in atomic units.}
\label{diplyA}
\end{figure}

Figure~\ref{Ly2all} shows the potential energy for the transitions that contribute to the profile of Lyman-$\alpha$ and Figure \ref{diplyA}
shows the electronic transition dipole moments for Lyman-$\alpha$ due to H perturbed by collisions with protons. 

Table~\ref{TabLyA} lists the wavelengths of the satellite lines with the upper and lower state identifications and the distance 
of the atom-ion pair at which the potential extreme occurs. Several of these features can be identified in the total Lyman-$\alpha$ 
profile of Figures \ref{LyA13} and \ref{LyA1012}. 

We list in Table~\ref{TabLyA} the satellite around $\lambda=1400$~\AA, 
observed by
\citet{Koester1985} and \citet{Nelan1985}, 
who proposed the lines are caused by quasi-molecular 
absorption of the H$_{2}^{+}$ molecule.

\begin{table}
\centering 
  \caption{Allowed transitions and satellites due to H - H$^{+}$ collisions of Lyman-$\alpha$ to the distance R(\AA)
of the atom-ion pair at which the potential extreme occurs.}
\begin{tabular}{|c |  c | c | c | c | c |} 
\hline
Label & Upper & Lower & $\Delta\nu_{H_{2}^{+}}$ & $\lambda_{H_{2}^{+}}$ & $R$\\
      & level & level & 	$(cm^{-1})$	&	$(\AA)$ 	& $(\AA)$	\\
\hline
1 & $2p\pi_{u}$ & $1s\sigma_{g}$ &	-1628.54 &	1240.56 & 4.72\\
2 & $3p\sigma_{u}$ & $1s\sigma_{g}$ & - & - & -\\
3 & $4f\sigma_{u}$ & $1s\sigma_{g}$ & -1172.40 &	1233.58 & 11.07\\
4 & $2s\sigma_{g}$ & $ 2p\sigma_{u} $ & 8139.91 & 1106.47 & 2.65 \\
5 & $3d\sigma_{g}$ & $ 2p\sigma_{u} $ &-11055.90 &	1404.87 & 4.56\\
  &                &                  &  16001.33  & 1017.93 & 1.39 \\
6 & $3d\pi_{g}$ & $ 2p\sigma_{u} $ & 21384.28 &	965.05 & 1.67\\
\hline
\end{tabular} 
\label{TabLyA}
\end{table}

\subsubsection{Lyman-$\beta$}

The profile of Lyman-$\beta$ depends on 10 transitions. Table \ref{TabLyB} lists all the wavelengths of these transitions and the possible lines satellites.
Satellites of Lyman-$\beta$ were observed in the spectra of the DA white dwarf Wolf~1346 with the Hopkins Ultraviolet Telescope (HUT) 
\citep{Koester1996}.
\citet{Koester1998} observed 
satellites absorption features at $ \lambda=1060$~\AA~ and $ \lambda=1080$~\AA
in the spectra of 4 other targets obtained
with the Orbiting and Retrievable Far and Extreme Ultraviolet Spectrometer (ORFEUS).
These satellites are
due to the contributions of the transitions $4f\pi_{u}$-$1s\sigma_{g}$ and $5g\sigma_{g}$-$ 2p\sigma_{u} $  of the profile of Lyman-$\beta$, 
according to the profiles of calculated by \citet{Allard1998a}.

\begin{figure}
\centering 
\includegraphics[width=84mm,clip=]{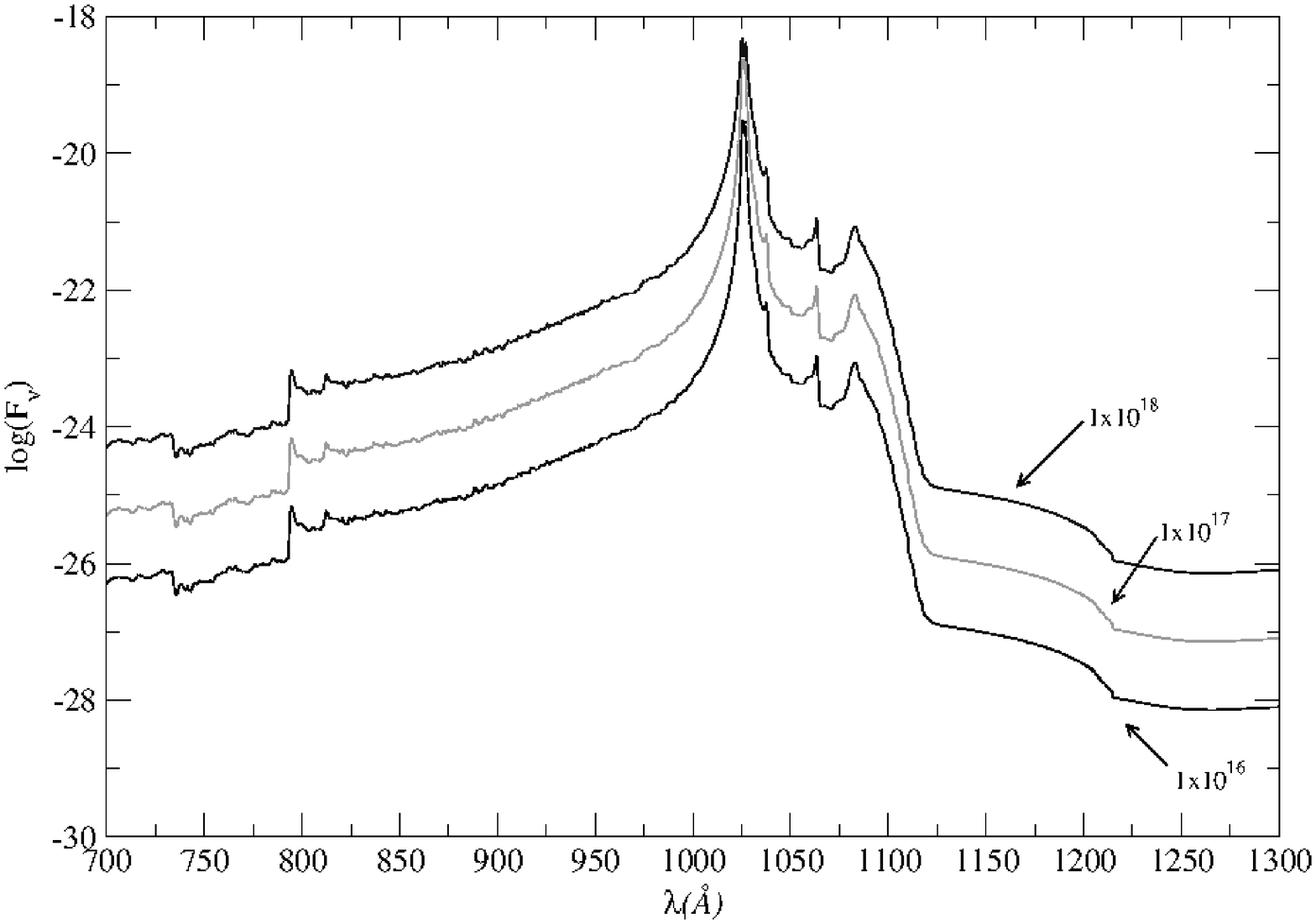}
\caption{Lyman-$\beta$ profile for 
T$=4\,000$~K and
density of H$^{+}$ perturbers of
$n_{H}^{+}=10^{16}$, $n_{H}^{+}=10^{17}$ and $n_{H}^{+}=10^{18}$ cm$^{-3}$ (from bottom to top).}
\label{LyB13}
\end{figure}

\begin{figure}
\centering 
\includegraphics[width=84mm,clip=]{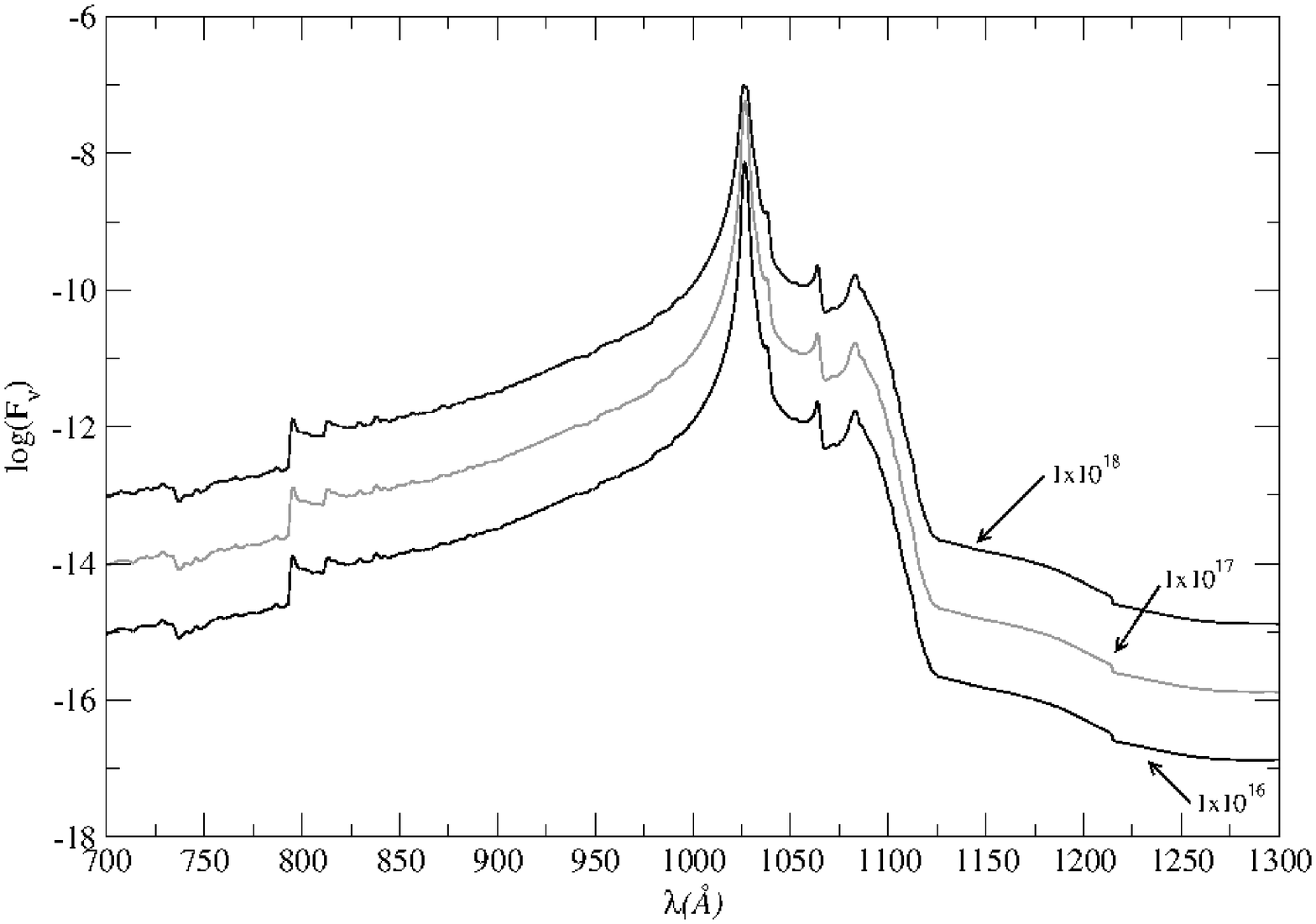}
\caption{Lyman-$\beta$ profile for
T$=12\,000$~K, and 
density of H$^{+}$ perturbers of
$n_{H}^{+}=10^{16}$, $n_{H}^{+}=10^{17}$ and $n_{H}^{+}=10^{18}$ cm$^{-3}$ (from bottom to top). }
\label{LyB1012}
\end{figure}

\begin{figure}
\centering 
\includegraphics[width=84mm,clip=]{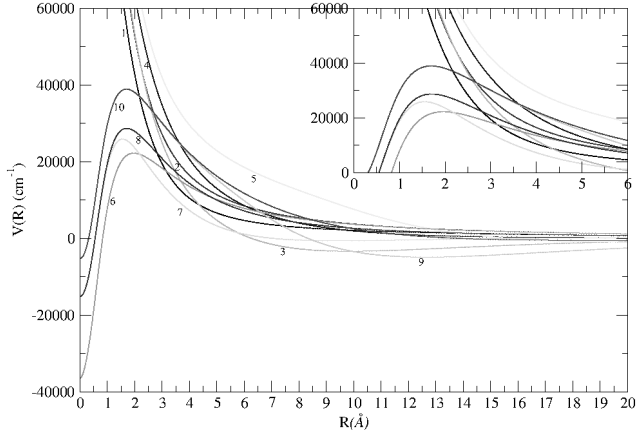}
\caption{Transition potential energies as a function of interatomic distance of the H$_{2}^{+}$ of Lyman-$\beta$.}
\label{Ly3all}
\end{figure}

\begin{figure}
\centering 
\includegraphics[width=84mm,clip=]{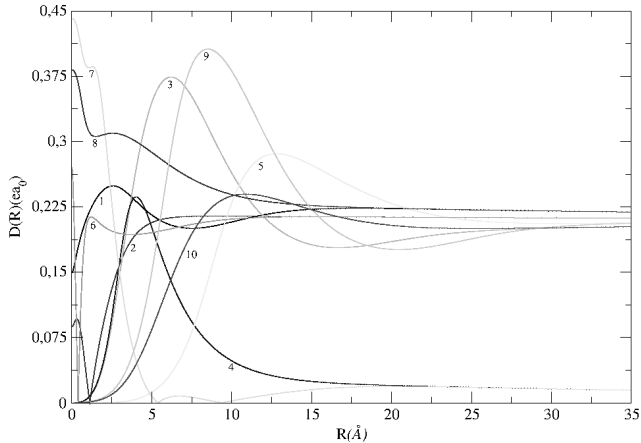}
\caption{Electronic transition dipole moments of Lyman-$\beta$ due to H perturbed by collisions with protons. 
The $D(R)$ is given here in atomic units.}
\label{diplyB}
\end{figure}

The Ly$\beta$ line profiles shown in Figuress \ref{LyB13} and \ref{LyB1012} are
for temperatures of $4\,000$~K and $12\,000$~K,  and densities of protons $10^{16}$, $10^{17}$ and $10^{18}$ cm$^{-3}$.
Figure \ref{Ly3all} shows all the differences of the potential energy of the transitions that contribute to the 
Lyman-$\beta$ line profile.
In Figure \ref{diplyB}, we plot the dipole D(R) for transitions of the Lyman-$\beta$ due to H perturbed by collisions with protons. 

\begin{table}
\centering 
\caption{Allowed transitions and satellites due to H - H$^{+}$ collisions of Lyman-$\beta$ to the distance R(\AA)
of the atom-ion pair at which the potential extreme occurs.} 
\begin{tabular}{|c |  c | c | c | c | c |} 
\hline
Label & Upper & Lower & $\Delta\nu_{H_{2}^{+}}$ & $\lambda_{H_{2}^{+}}$ & $R$\\
      & level & level & 	$(cm^{-1})$	&	$(\AA)$ 	& $(\AA)$	\\
\hline
1 & $3p\pi_{u}$ & $1s\sigma_{g}$ & - & - & -  \\
2 & $4p\sigma_{u}$ & $1s\sigma_{g}$ &	- &	- & - \\
3 & $4f\pi_{u}$ & $1s\sigma_{g}$ &	-3361.25 & 1062.64 & 9.85 \\
4 & $5f\sigma_{u}$ & $1s\sigma_{g}$ & - & - & - \\
5 & $6h\sigma_{u}$ & $1s\sigma_{g}$ &	-1037.33 &	1037.03 & 21.45 \\
6 & $3s\sigma_{g}$ & $ 2p\sigma_{u} $ &	22200.04 & 835.65 & 1.96 \\
7 & $4d\sigma_{g}$ & $ 2p\sigma_{u} $ &	-633.34 & 1032.71 & 9.46 \\
  &                &                  & 25868.59   & 810.80 & 1.55 \\           
8 & $4d\pi_{g}$ & $ 2p\sigma_{u} $ & 28666.11 & 792.82 & 1.70 \\
9 & $5g\sigma_{g}$ & $ 2p\sigma_{u} $ & -4899.9272 & 1080.31 & 12.65 \\
  &                &                  &  38880.59  & 733.42 & 1.69 \\ 
10 & $5g\pi_{g}$ & $ 2p\sigma_{u} $ & -516.61 &	1031.46 & 18.89 \\
  &                &                  &  38914.86 & 733.24 & 1.69 \\
\hline
\end{tabular}
\label{TabLyB}
\end{table}

\subsubsection{Lyman-$\gamma$}

There are 14 allowed transitions due to collisions of H - H$^{+}$ and the satellites 
predicted to occur are listed in Table~\ref{TabLyG}.
The satellite absorption features at $ \lambda=995$~\AA~ was observed in a
HUT spectrum and later in spectra recorded by ORFEUS. This Lyman-$\gamma$ satellite was also visible in the spectra of the white 
dwarf $CD-38^\circ10980$ \citep{Wolff2001} and 
Sirius B \citep{Holberg2003}. \citet{Allard2004} calculated the profiles of Lyman-$\gamma$ line based on accurate 
{\it ab initio} potentials to explain the observation of this feature.

\begin{table}
\centering 
\caption{Allowed transitions and satellites due to H - H$^{+}$ collisions of Lyman-$\gamma$ to the distance R(\AA)
of the atom-ion pair at which the potential extreme occurs.} 
\begin{tabular}{|c |  c | c | c | c | c |} 
\hline
Label & Upper & Lower & $\Delta\nu_{H_{2}^{+}}$ & $\lambda_{H_{2}^{+}}$ & $R$\\
      & level & level & 	$(cm^{-1})$	&	$(\AA)$ 	& $(\AA)$	\\
\hline
1 & $4p\pi_{u}$ & $1s\sigma_{g}$ &	- &	- & - \\
2 & $5p\sigma_{u}$ & $1s\sigma_{g}$ & - & - & - \\
3 & $5f\pi_{u}$ & $1s\sigma_{g}$ &	-192.78 & 974.62 & 16.65 \\
4 & $6f\sigma_{u}$ & $1s\sigma_{g}$ & - & - & - \\
5 & $6h\pi_{u}$ & $1s\sigma_{g}$ & -2188.78 & 993.96 & 20.69 \\
6 & $7h\sigma_{u}$ & $1s\sigma_{g}$ & -230.90 & 974.98 & 29.66 \\
7 & $8j\sigma_{u}$ & $1s\sigma_{g}$ & -784.17 & 980.27 & 36.06 \\
8 & $4s\sigma_{g}$ & $ 2p\sigma_{u} $ &	30153.49 & 752.16 & 1.82 \\
9 & $5d\sigma_{g}$ & $ 2p\sigma_{u} $ &	32004.30 & 741.83 & 1.62 \\
10 & $5d\pi_{g}$ & $ 2p\sigma_{u} $ & 33520.98 &	733.58 & 1.70 \\
11 & $6g\sigma_{g}$ & $ 2p\sigma_{u} $ &	-1395.38 &	986.18 & 17.79 \\
  &                &                  &	39005.56 &	705.21 & 1.69 \\
12 & $6g\pi_{g}$ & $ 2p\sigma_{u} $ & 39025.78 & 705.11 & 1.69 \\
13 & $7i\sigma_{g}$ & $ 2p\sigma_{u} $ &	-2627.07 &	998.31 & 25.06 \\
  &                &                  &	42340.74 &	689.00 & 1.70 \\
14 & $7i\pi_{g}$ & $ 2p\sigma_{u} $ & -565.52 &	978.18 & 31.65 \\
  &                &                  &	42342.54 &	688.99 & 1.70 \\
\hline
\end{tabular}
\label{TabLyG}
\end{table}

\begin{figure}
\centering 
\includegraphics[width=84mm,clip=]{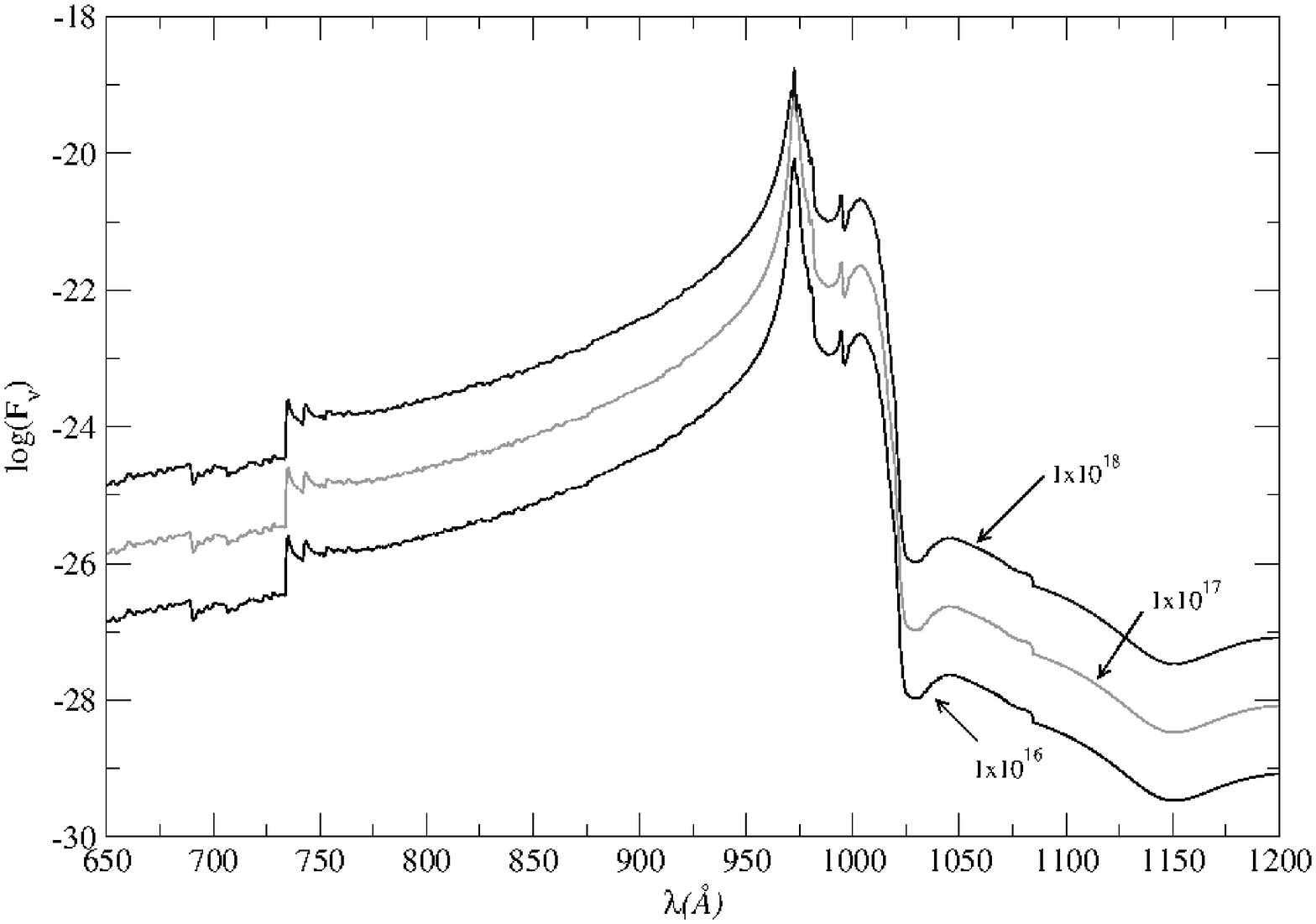}
\caption{Variation of the Lyman-$\gamma$ profile with density of H$^{+}$ perturbers (T$=4\,000$~K), $n_{H}^{+}=10^{16}$, $n_{H}^{+}=10^{17}$
 and $n_{H}^{+}=10^{18}$ cm$^{-3}$ (from bottom to top).}
\label{LyG13}
\end{figure}

\begin{figure}
\centering 
\includegraphics[width=84mm,clip=]{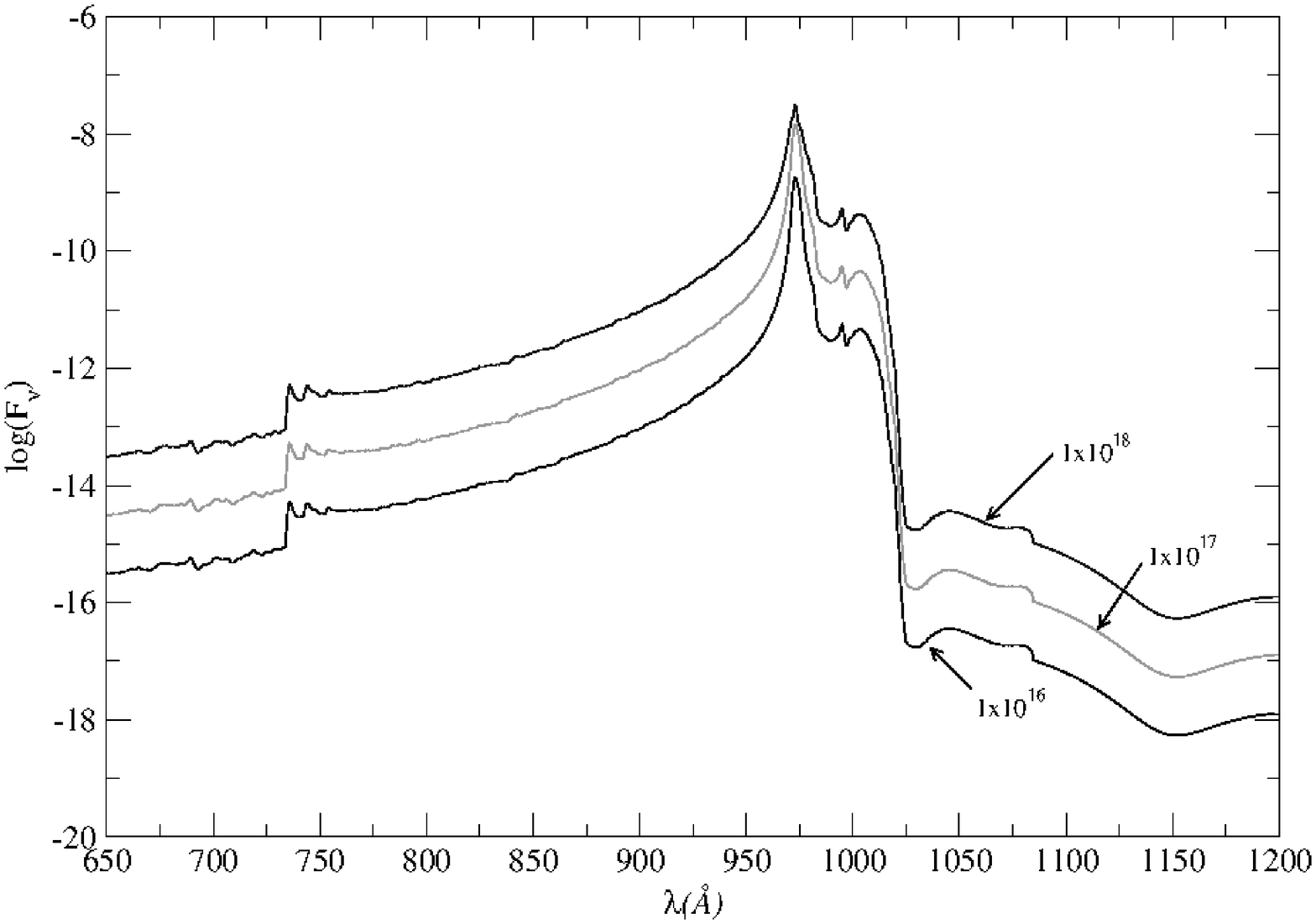}
\caption{Variation of the Lyman-$\gamma$ profile with density of H$^{+}$ perturbers (T$=12\,000$~K), $n_{H}^{+}=10^{16}$, 
$n_{H}^{+}=10^{17}$ and $n_{H}^{+}=10^{18}$ cm$^{-3}$ (from bottom to top).}
\label{LyG1012}
\end{figure}

The line profile calculations shown in Figure~\ref{LyG13} and Figure~\ref{LyG1012} have been 
evaluated at a temperature of $4\,000$~K and $12\,000$~K for differents densities of protons $10^{16}$, $10^{17}$ and $10^{18}$ cm$^{-3}$.

\begin{figure}
\centering 
\includegraphics[width=84mm,clip=]{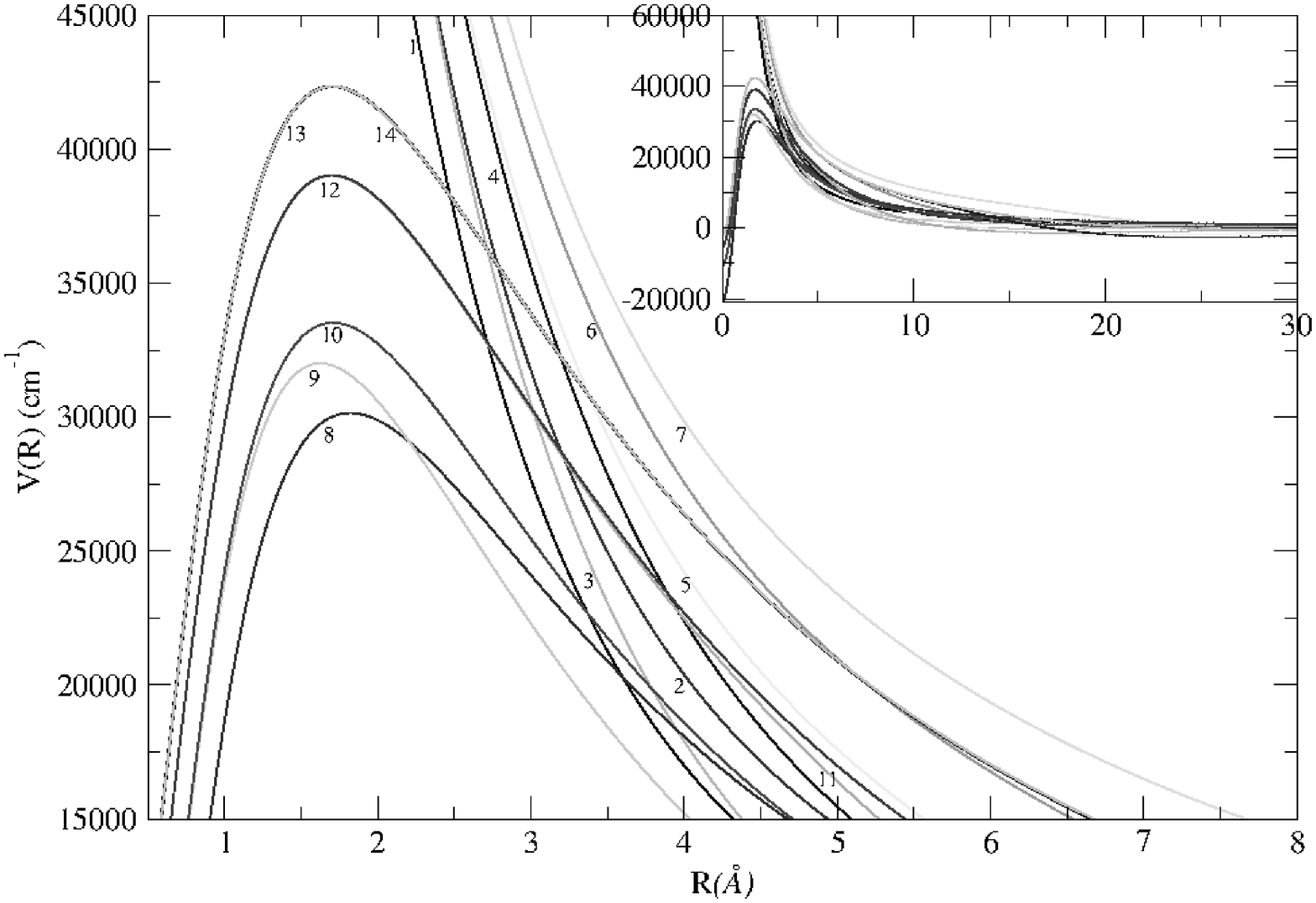}
\caption{
Transition potential energy differences
as a function of interatomic distance of the H$_{2}^{+}$ of Lyman-$\gamma$.}
\label{Ly4all}
\end{figure}

\begin{figure}
\centering 
\includegraphics[width=84mm,clip=]{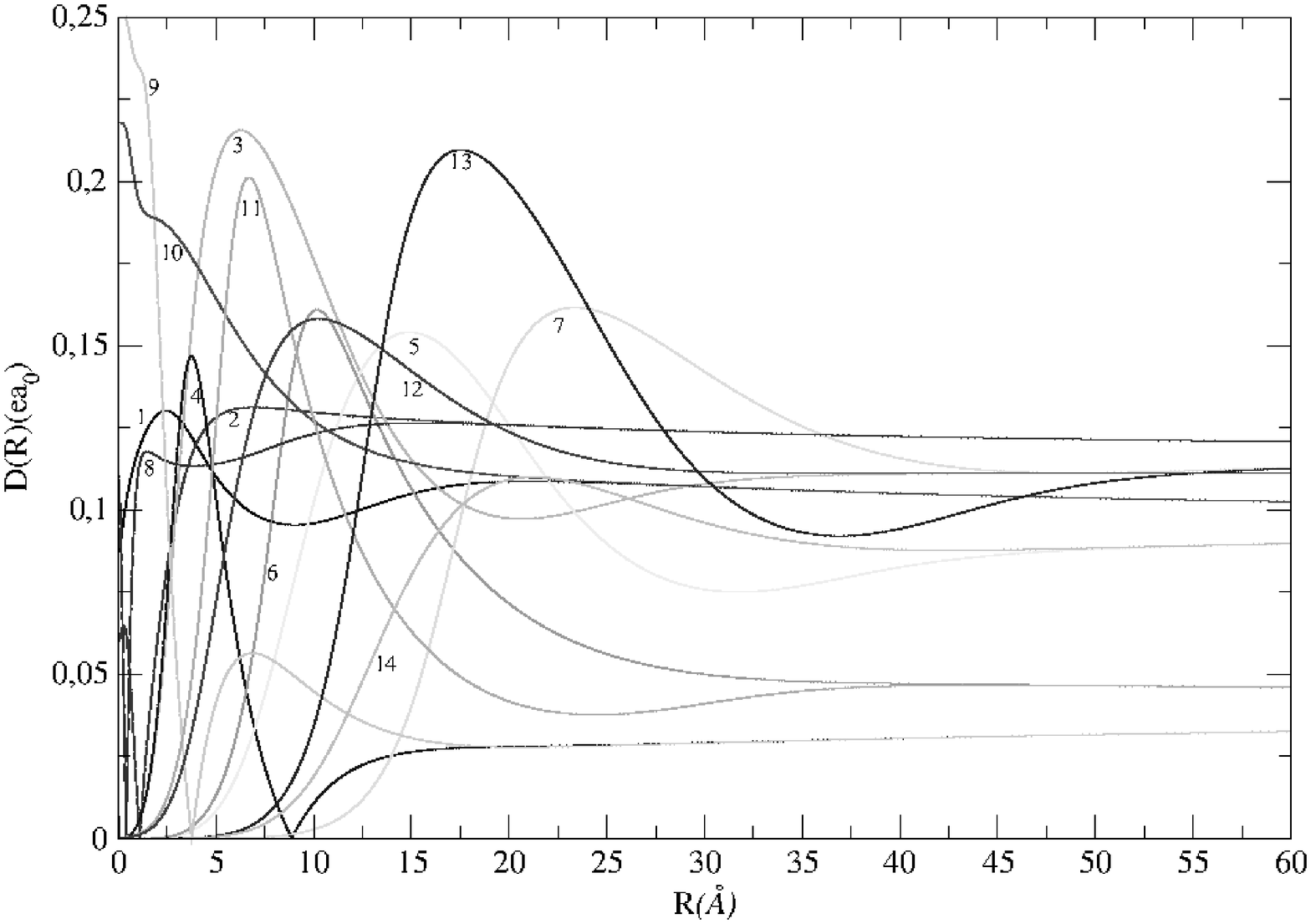}
\caption{Electronic transition dipole moments of Lyman-$\gamma$ due to H perturbed by collisions with protons. 
The $D(R)$ is given here in atomic units.}
\label{diplyG}
\end{figure}

Figure \ref{Ly4all} shows all the potential energy differences for the 
transitions that contribute to the Lyman-$\gamma$ line profile and Figure \ref{diplyG}
plots the electronic transition dipole moments for Lyman-$\gamma$ due to H perturbed by collisions with protons.

\subsubsection{Lyman-$\delta$}

There are 18 allowed transitions due to collisions of H and H$^{+}$, and the satellites predicted to occur are listed in Table~\ref{TabLyD}.
\begin{table}
\centering 
\caption{Allowed transitions and satellites due to H - H$^{+}$ collisions of Lyman-$\delta$ to the distance R(\AA)
of the atom-ion pair at which the potential extreme occurs.}
\begin{tabular}{|c |  c | c | c | c | c |} 
\hline
Label & Upper & Lower & $\Delta\nu_{H_{2}^{+}}$ & $\lambda_{H_{2}^{+}}$ & $R$\\
      & level & level & 	$(cm^{-1})$	&	$(\AA)$ 	& $(\AA)$	\\
\hline
1 & $5p\pi_{u}$ & $1s\sigma_{g}$ & - & - & - \\
2 & $6p\sigma_{u}$ & $1s\sigma_{g}$ & - & - & - \\
3 & $6f\pi_{u}$ & $1s\sigma_{g}$ & - & - & - \\
4 & $7f\sigma_{u}$ & $1s\sigma_{g}$ & - & - & - \\
5 & $7h\pi_{u}$ & $1s\sigma_{g}$ &	-679.23 & 956.16 & 27.56 \\
6 & $8h\sigma_{u}$ & $1s\sigma_{g}$ & - & - & - \\
7 & $8j\pi_{u}$ & $1s\sigma_{g}$ & -1413.57 & 962.93 & 36.17 \\
8 & $9j\sigma_{u}$ & $1s\sigma_{g}$ & -312.56 & 952.82 & 44.76 \\
9 & $10l\sigma_{u}$ & $1s\sigma_{g}$ &	-582.87 & 955.28 & 54.98 \\
10 & $5s\sigma_{g}$ & $ 2p\sigma_{u} $ &	- &	- & - \\
11 & $6d\sigma_{g}$ & $ 2p\sigma_{u} $ &	- &	- & - \\
12 & $6d\pi_{g}$ & $ 2p\sigma_{u} $ & - & - & - \\
13 & $7g\sigma_{g}$ & $ 2p\sigma_{u} $ &	-76.95 & 950.69 & 26.02\\
14 & $7g\pi_{g}$ & $ 2p\sigma_{u} $ & - & - & - \\
15 & $8i\sigma_{g}$ & $ 2p\sigma_{u} $ &	-1119.44 &	960.21 & 31.54\\
16 & $8i\pi_{g}$ & $ 2p\sigma_{u} $ & -99.84 & 950.90 & 42.38 \\
17 & $9k\sigma_{g}$ & $ 2p\sigma_{u} $ &	-1589.40 &	964.56 & 41.91 \\
18 & $9k\pi_{g}$ & $ 2p\sigma_{u} $ & -475.45 &	954.31 & 49.00 \\
\hline
\end{tabular}  
\label{TabLyD}
\end{table}

\begin{figure}
\centering 
\includegraphics[width=84mm,clip=]{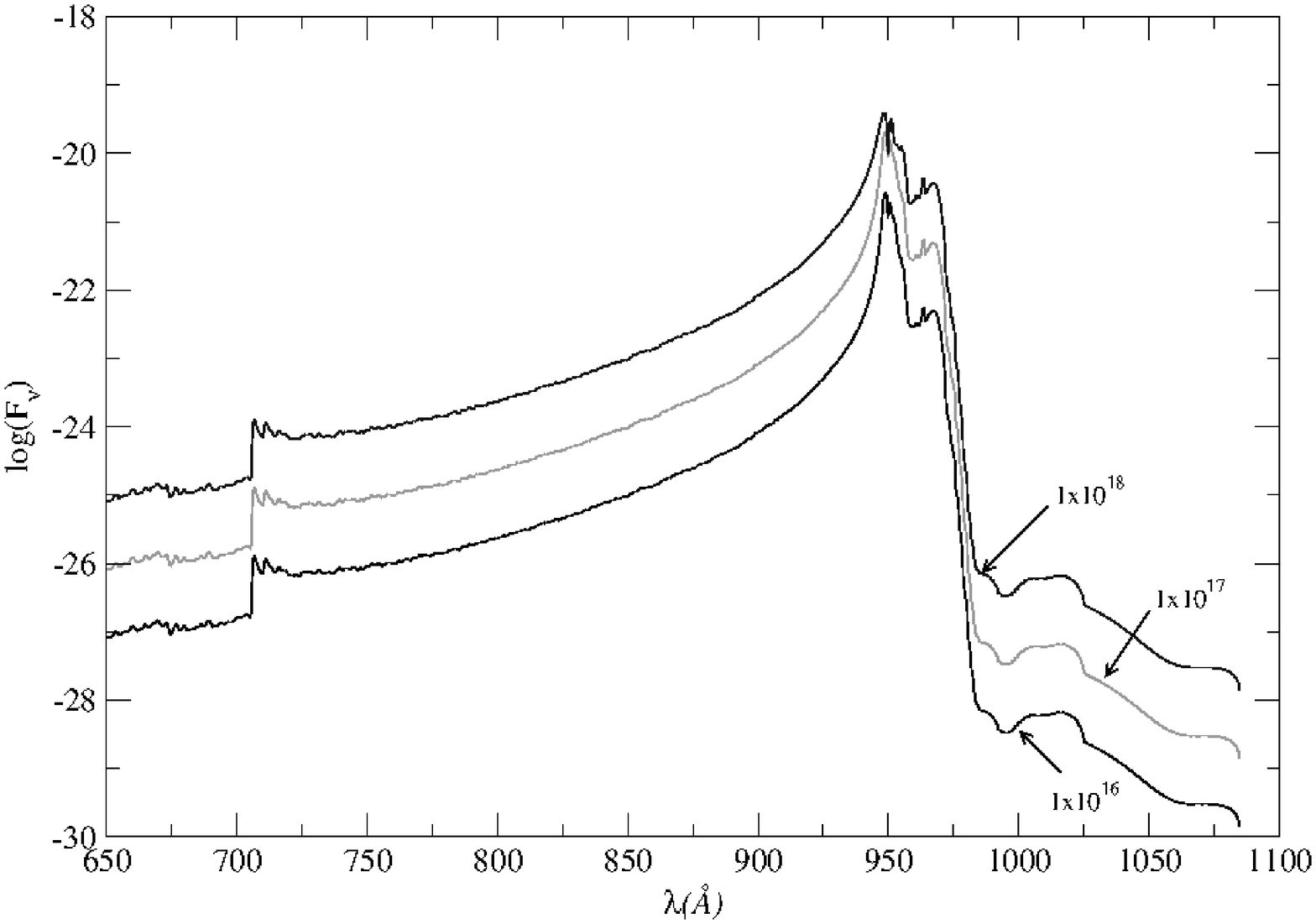}
\caption{Lyman-$\delta$ profile for density of H$^{+}$ perturbers (T$=4\,000$~K), 
$n_{H}^{+}=10^{16}$, $n_{H}^{+}=10^{17}$ and $n_{H}^{+}=10^{18}$ cm$^{-3}$  (from bottom to top).}
\label{LyD13}
\end{figure}

\begin{figure}
\centering 
\includegraphics[width=84mm,clip=]{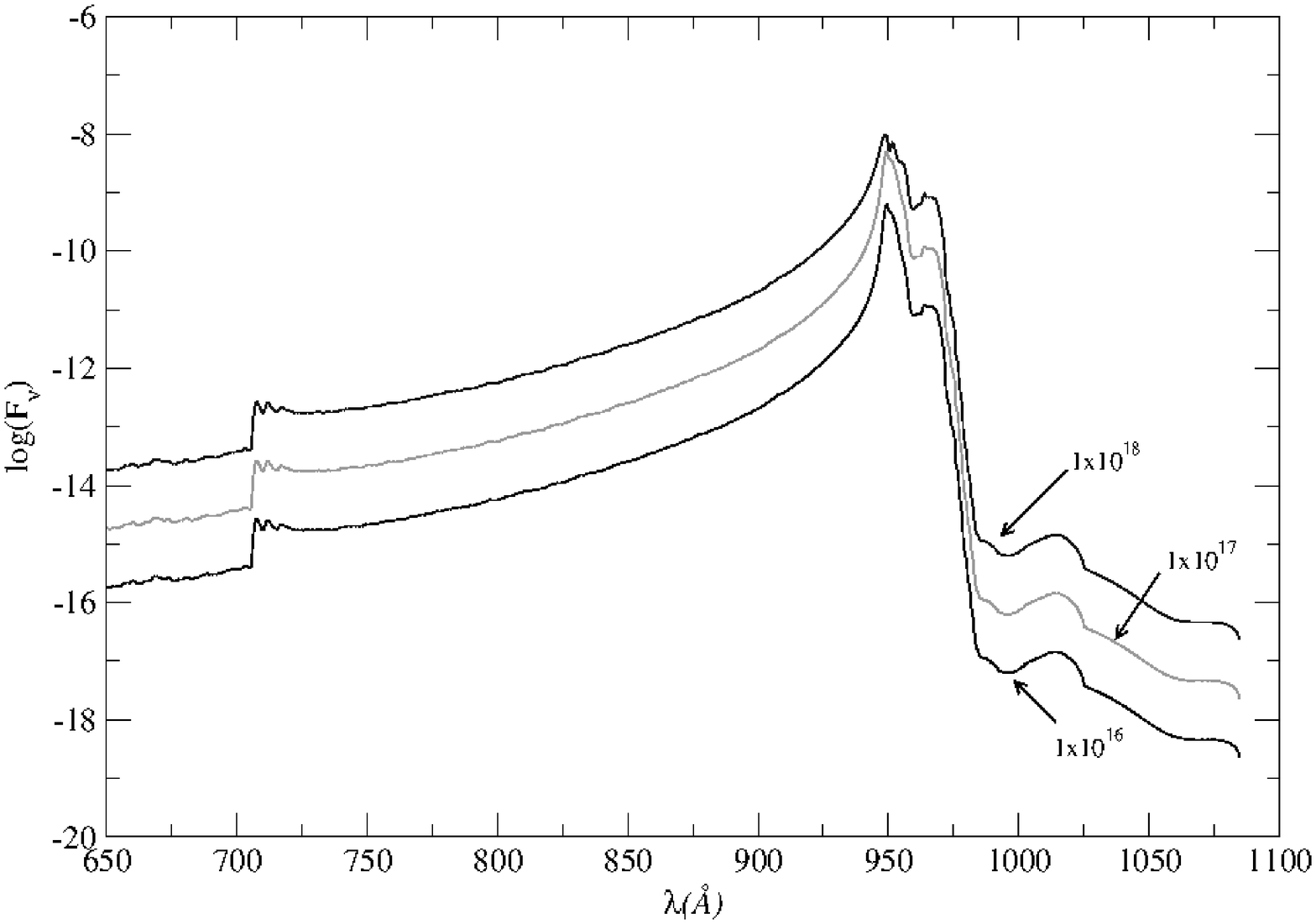}
\caption{Lyman-$\delta$ profile for density of H$^{+}$ perturbers (T$=12\,000$~K), $n_{H}^{+}=10^{16}$, $n_{H}^{+}=10^{17}$ and 
$n_{H}^{+}=10^{18}$ cm$^{-3}$  (from bottom to top).}
\label{LyD1012}
\end{figure}

Figures~\ref{LyD13} and \ref{LyD1012} show the line profiles as a function of temperature and density of perturbers.

\begin{figure}
\centering 
\includegraphics[width=84mm,clip=]{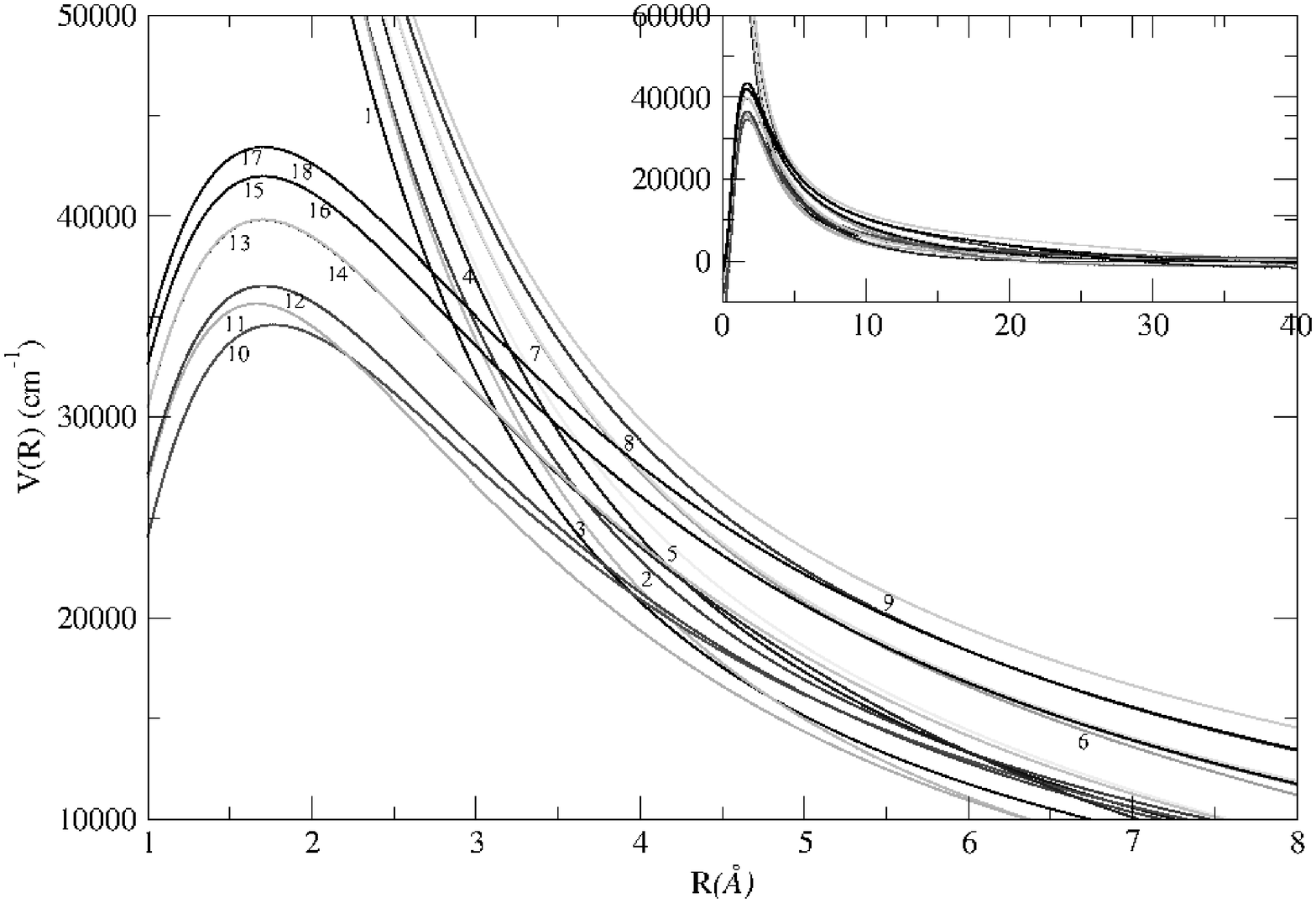}
\caption{
Potential energy differences for the allowed 
transitions,
as a function of interatomic distance of the H$_{2}^{+}$ of Lyman-$\delta$.}
\label{Ly5all}
\end{figure}

\begin{figure}
\centering 
\includegraphics[width=84mm,clip=]{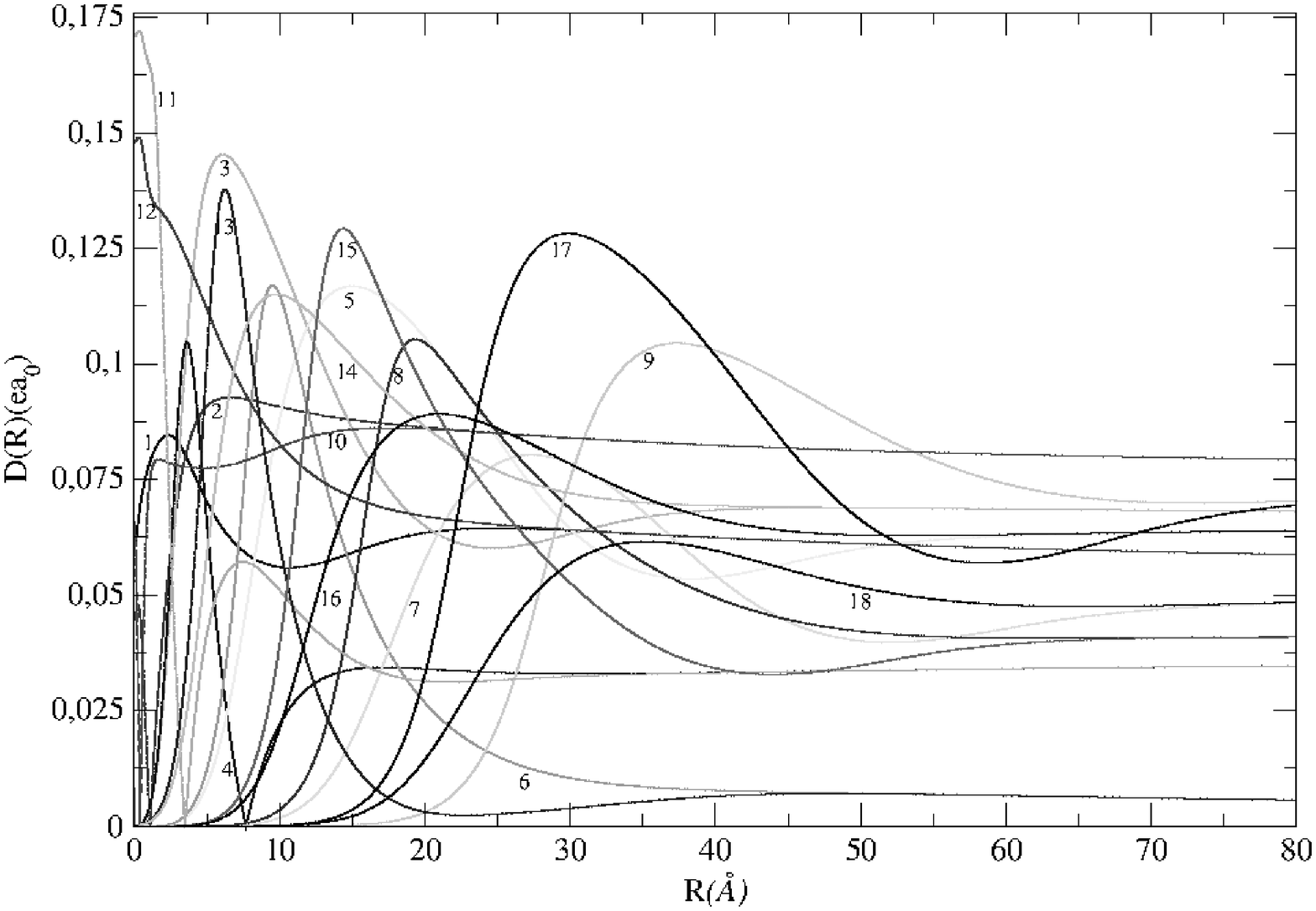}
\caption{Electronic transition dipole moments of Lyman-$\delta$ due to H perturbed by collisions with protons. 
The $D(R)$ is given here in atomic units.}
\label{diplyD}
\end{figure}

\begin{figure}
\centering 
\includegraphics[width=84mm,clip=]{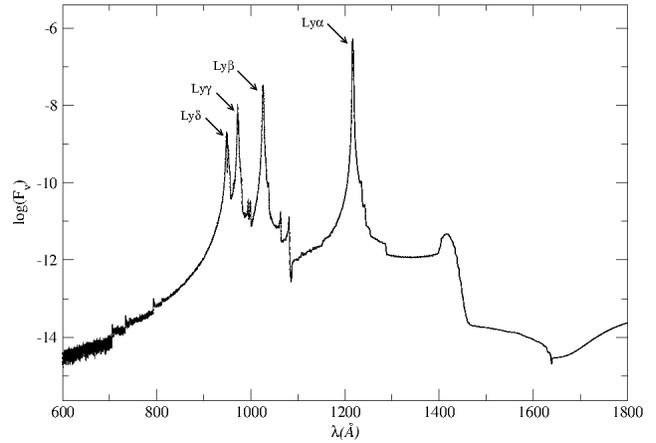}
\caption{Total profile of Lyman-$\alpha$, Lyman-$\beta$, Lyman-$\gamma$ and Lyman-$\delta$ perturbed by protons 
($n_{H}^{+}=10^{17}$ cm$^{-3}$) at a temperature T$=12\,000$~K.}
\label{LyALL11}
\end{figure}

Figure \ref{Ly5all} shows all the potential energy differences for the 
of the transitions that contribute to the line profile of Lyman-$\delta$. Figure \ref{diplyD}  shows the electronic transition dipole moments for Lyman-$\delta$ of the H$_{2}^{+}$ 
and Figure \ref{LyALL11} 
shows the total line profiles of Ly$\alpha$, Ly$\beta$, Ly$\gamma$ and Ly$\delta$ perturbed by protons for
a single T$=12\,000$~K and $n_{H}^{+}=10^{17}\,\mathrm{cm}^{-3}$. This figure is similar to Figure 3 obtained by \citet{Hebrard2003}, 
showing that our calculations and approximations are consistent with data already published.

\subsection{Balmer series}

The Balmer lines have transitions that must obey the same golden rules as the Lyman series transitions.

\subsubsection{Balmer-$\alpha$}

The total profile of Balmer-$\alpha$ depends on the 32 individual transitions. 
The line profile calculations shown in Figure \ref{painel} have been calculated at a single temperature of $12\,000$~K and protons density of $10^{17}$ cm$^{-3}$.

\begin{figure}
\centering 
\includegraphics[width=84mm,clip=]{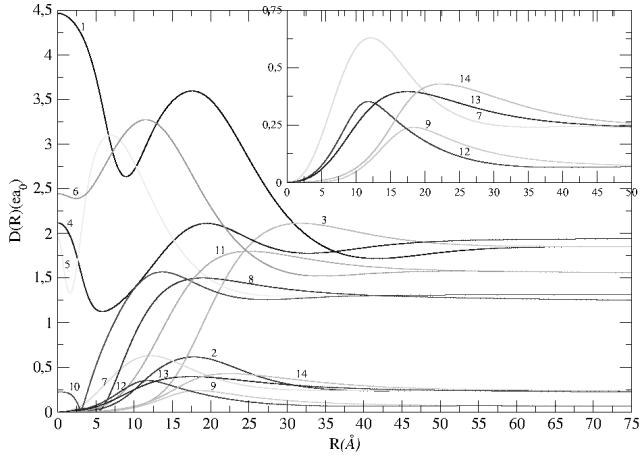}
\caption{Transition dipole moments for the satellites of Balmer-$\alpha$ due to H perturbed by collision with protons. 
The transitions plotted here are listed in Table \ref{TabBaA}. The $D(R)$ is given here in atomic units.}
\label{dipBaA}
\end{figure}

Table \ref{TabBaA} lists only the transitions that generate possible satellite lines on the wings of the Balmer line profiles. 
In Figure \ref{dipBaA} we plot the transition dipole moments for the satellites listed in Table \ref{TabBaA}.
In laboratory observations, measurements of the spectrum of a laser-produced plasma confirm that the $\lambda=8650$~\AA~ 
satellite exists in the H$\alpha$ red wing \citep{Kielkopf2002} due the transitions $5g\sigma_{g}$-$4f\sigma_{u}$ and 
$4f\pi_{u}$-$ 3d\pi_{g} $.

\begin{table}
\centering
\caption{Satellites due to H - H$^{+}$ collisions of Balmer-$\alpha$ to the distance R(\AA)
of the atom-ion pair at which the potential extreme occurs.} 
\begin{tabular}{|c |  c | c | c | c | c |} 
\hline
Label & Upper & Lower & $\Delta\nu_{H_{2}^{+}}$ & $\lambda_{H_{2}^{+}}$ & $R$\\
      & level & level & 	$(cm^{-1})$	&	$(\AA)$ 	& $(\AA)$	\\
\hline
1 & $5g\sigma_{g}$ & $4f\sigma_{u}$ & -3844.18 & 8783.59 & 13.10 \\
  &             &                & 4956.58 & 4954.01 & 5.60 \\
2 & $5g\sigma_{g}$ & $2p\pi_{u}$ &	-4996.57 &	9772.81 & 12.66\\
3 & $6h\sigma_{u}$ & $3d\sigma_{g}$ & 40925.21 & 1780.80 & 3.17\\
  &             &                & -702.54 & 6883.97 & 23.03 \\ 
4 & $4f\pi_{u}$ & $ 3d\sigma_{g} $ & -1271.27 & 7164.46 & 13.13 \\
  &             &                & 23797.73 & 2562.34 & 3.06 \\
5 & $4d\sigma_{g}$ & $ 3p\sigma_{u} $ &	-3214.16 &	8323.01 & 6.47 \\
  &             &                & 12145.77 & 3652.99 & 1.01 \\
6 & $4f\pi_{u}$ & $ 3d\pi_{g} $ &	-3658.85 &	8642.89 & 9.33\\
  &             &                & 7163.74 & 4465.72 & 1.69 \\      
7 & $4f\pi_{u}$ & $2s\sigma_{g}$ &	-5245.35 &	10016.33 & 8.69\\      
  &             &                & 37537.98 & 1895.12 & 1.00 \\       
8 & $5f\sigma_{u}$ & $2s\sigma_{g}$ &	-424.84 &	6754.83 & 15.47 \\      
9 & $6h\sigma_{u}$ & $2s\sigma_{g}$ & -1500.75 & 7284.22 & 20.33 \\
10 & $4d\sigma_{g}$ & $ 2p\pi_{u} $ & -634.82 &	6852.02 & 10.20 \\        
11 & $5g\pi_{g}$ & $ 2p\pi_{u} $ &	-601.70 &	6836.51 & 18.74 \\      
12 & $5g\sigma_{g}$ & $ 3p\sigma_{u} $ &	-6042.60 &	10885.60 & 12.08 \\
   &            &                  & 23287.20 & 2596.30 & 1.18 \\
13 & $5g\pi_{g}$ & $ 3p\sigma_{u} $ & -1130.91 &	7093.14 & 16.58\\
   &            &                  & 23303.15 & 2595.23 & 1.18 \\ 
14 & $6h\sigma_{u}$ & $ 3d\pi_{g} $ & -1117.22 &	7086.25 & 21.38\\
   &            &                  & 23163.48 & 2604.67 & 2.10 \\
\hline
\end{tabular} 
\label{TabBaA}
\end{table}

\begin{figure}
\centering 
\includegraphics[width=84mm,clip=]{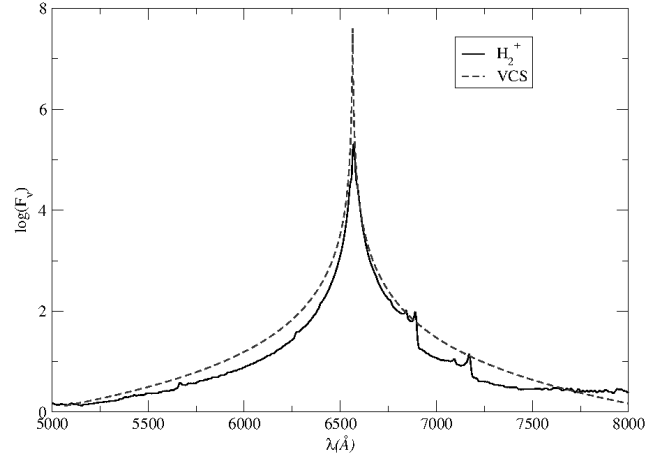}
\caption{Comparison between the profile of Balmer-$\alpha$ of H$_{2}^{+}$ with the profile VCS - electron Stark broadening, 
for T$=12\,000$~K and density $10^{16}$ cm$^{-3}$.}
\label{ckoester12K}
\end{figure}

Figure \ref{ckoester12K} shows the comparison between the VCS \citep{Vidal1973} profile calculated by Koester (private communication) for T$=12\,000$~K and 
$n_{H}^{+}$ density $10^{16}$cm$^{-3}$, with the profile of H$_{2}^{+}$ calculated for the same temperature and density.
It is observed that at high energies both profiles coincide asymptotically, while, 
due to possible formation of low-energy molecules, discrepancies arise between the profiles.

\subsubsection{Balmer-$\beta$}

\begin{figure}
\centering 
\includegraphics[width=84mm,clip=]{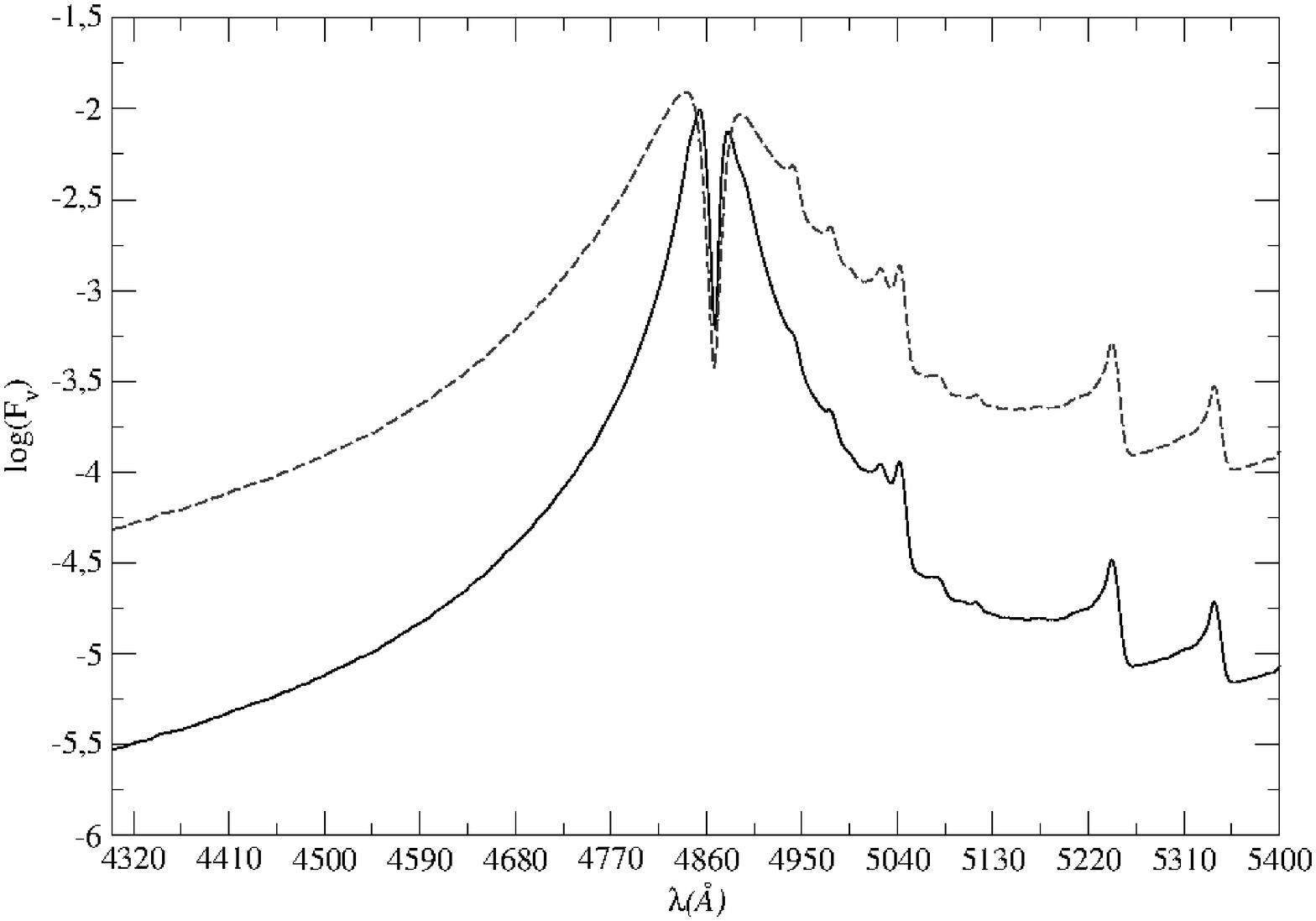}
\caption{Solid line is the Balmer-$\beta$ profile with density $n_{H}^{+}=9.324\times10^{16}$ cm$^{-3}$ of the perturbers and $T=13\,526$~K.   
Dashed line is the Balmer-$\beta$ profile with density $n_{H}^{+}=1.554\times10^{18}$ cm$^{-3}$ of the perturbers and $T=13\,526$~K.
The line profile can be used for diagnostic of the physical conditions of a laboratory measurements.}
\label{sandia12_13}
\end{figure}

The total profile of Balmer-$\beta$ depends on the 46 individual transitions. Figure \ref{sandia12_13} shows the profile of H$\beta$ 
calculated for protons density of $9.324\times10^{16}$ cm$^{-3}$ and $1.554\times10^{18}$ cm$^{-3}$ with $T=13\,526$~K. 
We can see the center of the line is split and asymmetric.

\begin{figure}
\centering 
\includegraphics[width=84mm,clip=]{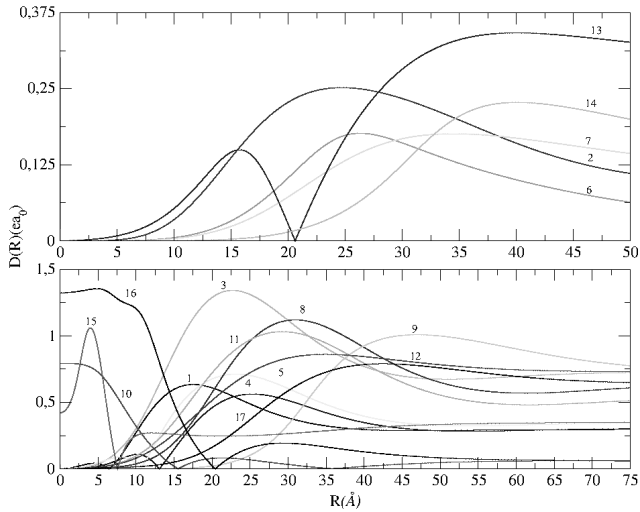}
\caption{Electronic Transition dipole moments for the satellites of Balmer-$\beta$ due to H perturbed by collision with protons. The transitions plotted here are listed in Table \ref{TabBaB}. The $D(R)$ is given here in atomic units.}
\label{dipBaB}
\end{figure}

Table \ref{TabBaB} lists the lines satellites due to the H$\beta$ and Figure \ref{dipBaB} plots the respective transitions listed in the same table.
The line profile calculations shown in Figure \ref{painel} is for temperature of $12\,000$~K 
and protons density of $10^{17}$ cm$^{-3}$.

\begin{table}
\centering
\caption{Satellites due to H - H$^{+}$ collisions of Balmer-$\beta$ to the distance R(\AA)
of the atom-ion pair at which the potential extreme occurs.} 
\begin{tabular}{|c |  c | c | c | c | c |} 
\hline
Label & Upper & Lower & $\Delta \nu_{H_{2}^{+}}$ & $\lambda_{H_{2}^{+}}$ & $R$\\
      & level & level & 	$(cm^{-1})$	&	$(\AA)$ 	& $(\AA)$	\\
\hline    
1 & $5f\pi_{u}$ & $2s\sigma_{g}$ & -992.00 & 5110.59 & 13.83 \\
2 & $6h\pi_{u}$ & $2s\sigma_{g}$ &	-2674.11 &	5591.24 & 19.90\\
3 & $5g\delta_{g}$ & $2p\pi_{u}$ & -1466.46 & 5237.59 & 16.80\\
4 & $6g\sigma_{g}$ & $ 2p\pi_{u} $ & -1483.16  & 5242.17 & 17.74 \\
5 & $6g\sigma_{g}$ & $ 3p\sigma_{u} $ &	-2040.74 &	5400.01 & 16.50\\
  &             &                & 23365.48 & 2276.62 & 1.19 \\   
6 & $7i\sigma_{g}$ & $3p\sigma_{u}$ &	-2977.17 &	5687.62 & 24.59\\      
  &             &                & 26648.74 & 2118.28 & 1.19 \\                   
7 & $7i\pi_{g}$ & $3p\sigma_{u}$ &	-816.65 &	5065.20 & 29.99 \\    
  &             &                & 26649.62 & 2118.24 & 1.19 \\      
8 & $6h\pi_{u}$ & $3d\sigma_{g}$ & -1811.98 & 5334.12 & 21.66 \\
  &             &                & 35621.24 & 1779.97 & 3.17 \\      
9 & $8j\sigma_{u}$ & $ 3d\sigma_{g} $ & -381.85 &	4956.05 & 29.99 \\        
   &            &                & 41130.19 & 1621.02 & 3.19 \\    
10 & $5f\pi_{u}$ & $ 3d\pi_{g} $ &	-285.48 &	4932.49 & 16.46 \\    
   &            &                & 12088.84 & 3062.96 & 1.91 \\    
11 & $6h\pi_{u}$ & $ 3d\pi_{g} $ &	-2269.96 &	5467.69 & 20.64 \\
   &            &                  & 17843.96 & 2603.95 & 2.11 \\
12 & $6h\delta_{u}$ & $ 3d\pi_{g} $ &	-308.92 &	4938.20 & 28.53 \\
   &            &                  & 17875.84 & 2601.79 & 2.13 \\ 
13 & $7h\sigma_{u}$ & $ 3d\pi_{g} $ & -302.86 &	4936.72 & 29.57\\
   &             &                  & 21106.24 & 2400.06 & 2.12 \\
14 & $8j\sigma_{u}$ & $ 3d\pi_{g} $ & -853.70 &	5074.72 & 36.03\\
   &            &                  & 23255.19 & 2282.35 & 2.15 \\               
15 & $5d\sigma_{g}$ & $ 4f\sigma_{u} $ & -11612.48 &	11177.27 & 1.95\\
   &            &                  & 3925.46 & 4084.18 & 8.17 \\                 
16 & $6g\sigma_{g}$ & $ 4f\sigma_{u} $ & -878.51 &	5081.12 & 20.00\\
   &            &                  & 6602.41 & 3681.66 & 6.08 \\             
17 & $7i\pi_{g}$ & $ 4f\sigma_{u} $ & -405.68 &	4961.91 & 29.99\\
   &            &                  & 12418.51 & 3032.34 & 6.81 \\            
\hline
\end{tabular} 
\label{TabBaB}
\end{table} 

\subsubsection{Balmer-$\gamma$}

The total profile of Balmer$-\gamma$ depends on the 60 individual transitions. 
The line profile shown in Figure \ref{painel} is for a temperature of $12\,000$~K and proton density $10^{17}$ cm$^{-3}$.

\subsubsection{Balmer-$\delta$}

The total profile of Balmer-$\delta$ depends on the 74 individual transitions. 
The line profile calculations of Figure \ref{painel} have been done at a temperature of $12\,000$~K 
for the protons density $10^{17}$ cm$^{-3}$.

\subsubsection{Balmer-$\epsilon$}

The total profile of Balmer-$\epsilon$ depends on the 88 individual transitions. 
The line profile calculations of Figure \ref{painel} have been done at a temperature of $12\,000$~K 
for the protons density $10^{17}$ cm$^{-3}$.

\subsubsection{Balmer-$8$}

The total profile of Balmer-$8$ depends on the 102  individual transitions. 
The line profile calculations of Figure \ref{painel} have been done at a temperature of $12\,000$~K 
for the protons density $10^{17}$ cm$^{-3}$.

\subsubsection{Balmer-$9$}

The total profile of Balmer-$9$ depends on the 116 individual transitions. 
The line profile calculations of Figure \ref{painel} have been done at a temperature of $12\,000$~K 
for the protons density $10^{17}$ cm$^{-3}$.

\subsubsection{Balmer-$10$}

\begin{figure}
\centering 
\includegraphics[width=84mm,clip=]{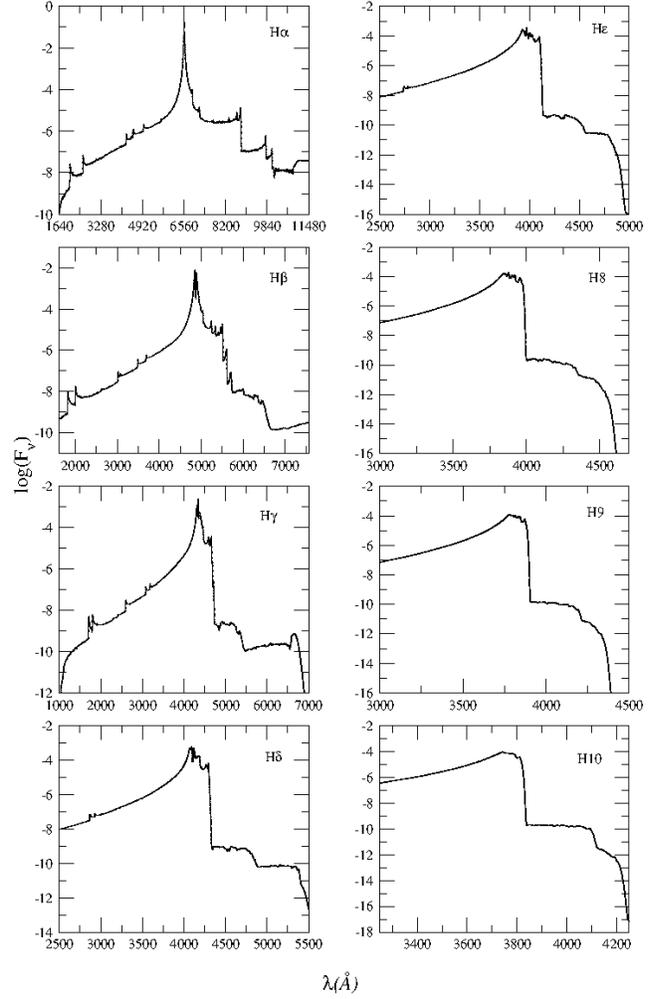}
\caption{Profiles of the Balmer series (H$\alpha$, H$\beta$, H$\gamma$, H$\delta$, H$\epsilon$, H8, H9, and H$10$) 
for ion density $n_{H}^{+}=10^{17}$ cm$^{-3}$ and T$=12\,000$~K.}
\label{painel}
\end{figure}

The total profile of Balmer-$10$ depends on the 130 individual transitions. 
The line profile calculations of Figure \ref{painel} have been done at a temperature of $12\,000$~K 
for the protons density $10^{17}$ cm$^{-3}$.

\begin{figure}
\centering 
\includegraphics[width=84mm,clip=]{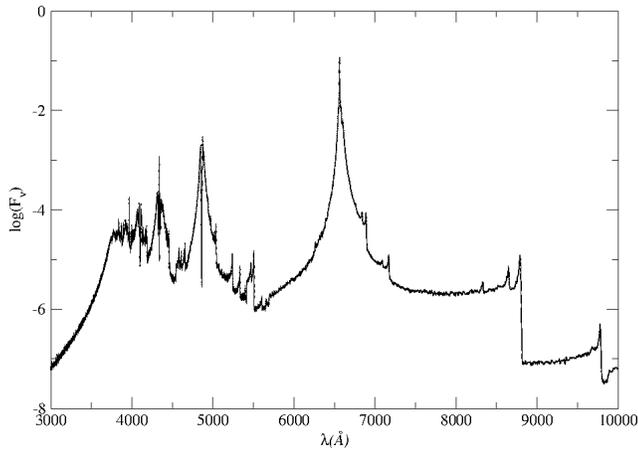}
\caption{Total profile of the Balmer for T$=12\,000$~K and $n_{H}^{+}=10^{17}$ cm$^{-3}$.}
\label{BaALL11}
\end{figure}

\section{Discussion and conclusion}

Satisfactory theory and data for the line profiles do exist due to the electron Stark broadening of neutral hydrogen \citep{Lemke1997, 
Vidal1973}, and for 21 optical lines of neutral helium \citep{Barnard1969, Beauchamp1997}. The first three Lyman lines of H broadened by 
ionized and neutral perturbers, including a number of satellite features, are well described by the work of Nicole Allard and 
collaborators \citep[e.g.][and papers cited therein]{Allard2009}, and the atomic data are obtained from a number of atomic databases, 
predominantly the line lists from Kurucz and collaborators \citep{Kurucz1995}, and the Vienna Atomic Line Database (VALD) \citep{Kupka1999, Kupka2000, 
Ryabchikova1997, Piskunov1995}.
For all other cases the situation is much less satisfactory, so there is a need for theories and data to explain the line profiles and 
satellite lines.

We extended the potentials dipole and line profile calculations to n=10 for the collision of H on H$^{+}$ perturbers important at high densities,
as observed in the atmosphere of white dwarf stars.

The laboratory observations of a laser-produced plasma confirm that satellites appear on Lyman$-\alpha$ due to collisions with neutral 
atoms and protons. This experimental confirmation of the theory supports the identification of these features in the Lyman$-\alpha$ 
spectra of white dwarf and $ \lambda $ Bootis stars.
In high temperature plasmas such as laser-produced plasmas, dielectronic recombination process is very important. This process strongly 
affects the ion abundances and produces dielectronic satellite lines. The satellite lines appear near the resonance line through 
spontaneous radiative decay from the autoionizing states. Satellite line emissions from highly ionized ions are often used for plasma 
diagnostics which are important in various kinds of plasmas, e.g., Tokamak, inertial fusion, astrophysical and plasma sources, etc 
\citep{Yamamoto2004}. 
The comparison of the Lyman$-\alpha$ wing with line shape models is a tool for determining neutral and proton densities in a hydrogenic 
plasma.
These experiments also confirm that the variation of the radiative dipole moment is an important factor in determining the far wing 
emission of Lyman$-\alpha$.
When $D(R)$ differs significantly from its asymptotic value at an R close to the region forming a satellite, the strength of the wing 
may be enhanced (or diminished) considerably.

In a collision-induced spectrum such as this one, the dipole moment may be more important than the potential in determining the shape of 
the satellite.
These many-body perturbations alter the far line wing, adding multiple satellites and producing a strong continuum from the vacuum 
ultraviolet to the ultraviolet and visible.

The present calculations are done in an adiabatic approximation using a rectilinear trajectory. This should affect slightly the shape of 
the satellite, although no significant differences are expected.

The sum of the profiles calculated of Ly$\alpha$, Ly$\beta$, Ly$\gamma$ and Ly$\delta$, are shown in Figure~\ref{LyALL11} and the sum of the
profiles of Balmer series in Figure~\ref{BaALL11} to T$=12\,000$~K and $n_{H}^{+}=10^{17}$ cm$^{-3}$. The ion collision profiles differs 
significantly from the electron Stark broadening, showing not only satellite lines but also significant wing values that affect the opacity.

The data is available in our web page at http://astro.if.ufrgs.br/marcios.

\end{document}